\documentclass[11pt,aps,nofootinbib,prd,aps,epsf,floats,axodraw,amsmath,amssymb,amsfonts]{revtex4}
\usepackage{amsmath, amssymb}
\newcommand{\mathsym}[1]{{}}

\usepackage{graphicx}
\usepackage{amsmath}
\usepackage{amssymb}
\usepackage{bm}
\newcommand{\be}{\begin{equation}}
\newcommand{\ee}{\end{equation}}
\newcommand{\bea}{\begin{eqnarray}}
\newcommand{\eea}{\end{eqnarray}}

\newcommand{\rem}[1]{}
\newsavebox{\PSLASH}
 \sbox{\PSLASH}{$p$\hspace{-1.8mm}/}
 
\renewcommand{\theequation}{\thesection.\arabic{equation}}
\newcounter{saveeqn}
\newcommand{\add}{\addtocounter{equation}{1}}
\newcommand{\alpheqn}{\setcounter{saveeqn}{\value{equation}}%
\setcounter{equation}{0}%
\renewcommand{\theequation}{\mbox{\thesection.\arabic{saveeqn}{\alph{equation}}}}}
\newcommand{\reseteqn}{\setcounter{equation}{\value{saveeqn}}%
\renewcommand{\theequation}{\thesection.\arabic{equation}}}

 \newsavebox{\notrightarrow}
 \sbox{\notrightarrow}{$\to$\hspace{-4mm}/}
 
 \newsavebox{\PARTIALSLASH}
 \sbox{\PARTIALSLASH}{$\partial$\hspace{-1.6mm}/}
 
 \newsavebox{\ASLASH}
 \sbox{\ASLASH}{$A$\hspace{-2.1mm}/}
 
 \newsavebox{\KSLASH}
 \sbox{\KSLASH}{$k$\hspace{-1.8mm}/}
 
 \newsavebox{\LSLASH}
 \sbox{\LSLASH}{$\ell$\hspace{-1.8mm}/}
 
 \newsavebox{\QSLASH}
 \sbox{\QSLASH}{$q$\hspace{-1.8mm}/}
 
 \newsavebox{\DSLASH}
 \sbox{\DSLASH}{$D$\hspace{-2.2mm}/}
 
 \newsavebox{\DbfSLASH}
 \sbox{\DbfSLASH}{${\mathbf D}$\hspace{-2.8mm}/}
 
 \newsavebox{\DELVECRIGHT}
 \sbox{\DELVECRIGHT}{$\stackrel{\rightarrow}{\partial}$}
 
 \newcommand{\blue}{\IfColor{\textCadetBlue}{}}
\newcommand{\black}{\IfColor{\textBlack}{}}
\newcommand{\red}{\IfColor{\textRed}{}}
\newcommand{\green}{\IfColor{\textOliveGreen}{}}
\newcommand{\lila}{\IfColor{\textRedViolet}{}}

\begin{document}
\begin{flushright}
 [hep-th/math-ph]
\end{flushright}
\title{Brownian Motion in the Hilbert Space of Quantum States and the Stochastically Emergent Lorentz Symmetry: A Fractal Geometric Approach from Wiener Process to Formulating Feynman's Path-Integral Measure for Relativistic Quantum Fields}

\author{Amir Abbass Varshovi}\email{ab.varshovi@sci.ui.ac.ir/amirabbassv@gmail.com/amirabbassv@ipm.ir}

\begin{abstract}
\textbf{Abstract\textbf{:}} This paper aims to provide a consistent, finite-valued, and mathematically well-defined reformulation of Feynman's path-integral measure for quantum fields obtained by studying the Wiener stochastic process in the infinite-dimensional Hilbert space of quantum states. This reformulation will undoubtedly have a crucial role in formulating quantum gravity within a mathematically well-defined framework.\footnote{This paper is followed by two separate works, cited as references \cite{varshovi E-H} and \cite{varshovi}, which establish the importance and capability of the reformulation to include gravitational effects.} In fact, the present study is fundamentally different from any previous research on the relationship between Feynman's path-integral and the Wiener stochastic process. In this research, we focus on the fact that the classic Wiener measure is no longer applicable in infinite-dimensional Hilbert spaces due to fundamental differences between displacements in low and extremely high dimensions. Thus, an analytic norm motivated by the role of the fractal functions in the Wilsonian renormalization approach is worked out to properly characterize Brownian motion in the Hilbert space of quantum states on a compact flat manifold. This norm, the so-called \emph{fractal norm}, pushes the rougher functions (physically the quantum states with higher energies) to the farther points of the Hilbert space until the fractal functions as the roughest ones are moved to infinity. Implementing the Wiener stochastic process with the fractal norm, results in a modified form of the Wiener measure called the \emph{Wiener fractal measure}, which is a well-defined measure for Feynman'spath-integral formulation of quantum fields. Wiener fractal measure has a complicated formula of non-local terms but produces the Klein-Gordon action at the first order of approximation. Using complex integrals to compensate for the removal of non-local terms appearing in higher orders of approximation, the Wiener fractal measure turns into a complex measure and generates Feynman's path-integral formulation of scalar quantum fields. This brings us to the main objective of this study. Finally, some various significant aspects of quantum field theory (such as renormalizability, RG flow, Wick rotation, regularization, etc.) are revisited by means of the analytical aspects of the Wiener fractal measure. \\

\noindent \textbf{Keywords\textbf{:}} Brownian Motion, Weierstrass Function, Fractality, Wiener Fractal Measure, Feynman's Path-Integral Measure, Emergent Lorentz Symmetry, RG Flow.
\end{abstract}

 \maketitle


\section{Introduction}
\label{introduction}

\par One of the most important properties of Brownian motion in finite-dimensional vector spaces is that its formulation, such as the Wiener measure, has to be symmetric with respect to motion in various directions. However, such symmetry will break down in infinite-dimensional Hilbert spaces. This symmetry breaking can be demonstrated with a variety of indications or theoretical defects. For example, the components of any random displacement in an infinite dimensional Hilbert space, as an element of the Hilbert space, must tend to zero in very high dimensions. This shows that movements in low and extremely high dimensions are not included in the same way in Brownian motion in Hilbert spaces. For another example, the standard formula of Wiener measure in such Hilbert spaces will include an infinite-dimensional volume form at any section of time, which is obviously ill-defined as a mathematical entity. This indicates that the mentioned formula must be modified in infinite-dimensional spaces to suppress or control the integration in higher dimensions by an appropriate renormalization process. As a conclusion, motions in low and extremely high dimensions cannot be included in the Wiener measure formula symmetrically.

\par This symmetry breaking has some objective effects, whose investigation will contribute to a deeper understanding of the problem. To see this let $\mathcal H = L^2(S^1)$, be the space of $L^2$-functions on $S^1$. Each element of $\mathcal H$, say $\psi(\theta)=\sum_{n\in \mathbb Z} a_n e^{in\theta}$, is in fact, a sequence of complex numbers $\{ a_n \}_{n \in \mathbb Z}$ with $\sum_n |a_n|^2 < \infty$. When $a_0$ is changed to $a_0+\Delta$, the function's value is changed everywhere by $\Delta$. However, when $a_n$, for $n\ne 0$, is changed by the same amount, the function's oscillation amplitude increases by $\Delta$ for frequency $n$. The bigger $|n|$ is, the more the function fluctuates. Therefore, in terms of mathematical analysis, there is no symmetry between these movements. Moreover, if we assume that $\mathcal H$ is the Hilbert space of quantum states on $S^1$, then the same reasoning holds physically. The quantum particle's energy remains unchanged in the first shift, but it varies proportional to $n^2$ in the second alteration. These instances show that Brownian motion in $\mathcal H$ is entirely asymmetric from both analytic and physical viewpoints.

\par A thorough analysis can demonstrate that an infinite sequence of movements in infinite-dimensional Hilbert space $\mathcal H$ can affect the class of smoothness of $\psi(\theta)$ by repeatedly going through the shift $a_n \to a_n + \Delta_n$ over and over for ever-larger $n$. At the end of this process, there exist functions that are nowhere differentiable and are effectively fixed points of a physical Brownian motion, the so-called \emph{fractal} or \emph{Weierstrass functions}. Nowhere differentiability can similarly be extended to higher dimensional manifolds. From the viewpoint of quantum physics, such states represent particles with infinite amounts of energy. In quantum field theory, they are located at the extreme limit of the RG flow in the Wilsonian renormalization procedure \cite{wilson1, wilson2}. Therefore, it appears that to accurately introduce the physical Brownian motion in the infinite-dimensional Hilbert space of quantum states, we need to construct a physically motivated norm that pushes the fractal functions to the infinities of the Hilbert space. To generate or discover the promised norm we must perform an extremely precise mathematical analysis of the analytic features of fractal functions. This inquiry comprises over half of the current study.

\par As we will demonstrate in the present research, studying fractals is the most important step in interpreting and figuring out the physical Brownian motion in the infinite-dimensional Hilbert space of quantum states, comprehending the mathematics of Feynman's path-integral formulation of quantum fields, and grasping the analytical core of the Wilsonian renormalization procedure of interacting quantum fields. This is because fractals, as the elements at infinity, are the most immobile elements of the infinite-dimensional Hilbert space of quantum states and must be preserved along a physical Brownian motion process. Aspects of complexity that don't have any counterparts in vector spaces with finite dimensions. However, as the most intricate components of the Hilbert space of quantum states, fractals, in their ideal analytic form, are physically impossible to occur in quantum systems due to their infinite amount of energy. Nevertheless, the theoretical study of fractals will provide an analytic criterion to measure the fractality of each element of the Hilbert space so that this measurement would be the most crucial step toward comprehending the Brownian motion of quantum particles in the Hilbert space of quantum states and deriving a well-defined formulation of Feynman's path-integral measure for quantum fields. 

\par Fractals are, in fact, one of the most common features of nature in both macro and micro scales. Indeed, the self-similarity of natural structures, which is preserved in scale transformation within some specific range of spatial length (or energy) of physical systems stems from a well-understood symmetry of quantum field theories, known as conformal invariance \cite{sornette, piet}. However, although this symmetry usually breaks at some scale of energies for interacting field theories, the fundamental mathematical entities involved within the formulation of quantum physics show some intrinsic features of self-similarity (scale invariance) and nowhere differentiability \cite{kroger, feynman, laskin}. These features can be related to both physical (observable) structures and unobservable entities such as quantum states or quantum fields. In fact, such properties of quantum fields are mostly figured out and utilized via the renormalization group flow through with both Wilson's and Bogoliubov's approaches wherein the high energy contributions are effectively referred to \cite{bogo, shirkov1, shirkov2, shirkov3, wilson1, wilson2}. 

\par As a mathematical object a fractal structure on a $D$-dimensional Cartesian space is basically defined upon the generalization of fractal curves. A fractal curve is generally defined by analytic concepts such as measure, dimension, length, or differentiability. Due to Mandelbrot \cite{mandelbrot} the fractal curves are distinguished by a measure theoretic point of view. Through this viewpoint a fractal curve in $\mathbb{R}^D$ is, in fact, a function $\frak{F}:I \to \mathbb{R}^D$, with $\text{dim}_{HB}(\frak{F}(I))> 1$, wherein $\text{dim}_{HB}$ is the Hausdorff-Besicovitch dimension and $I\subset \mathbb R$ is an interval. However, according to the geometric-base definition, a fractal curve is a continuous embedding of $I$ in $\mathbb{R}^D$ which each of its subarc longer than a point has an infinite length \cite{nottale1}. This point of view also leads to the definition of nowhere rectifiable curves which are effectively known as maps from $I$ to $\mathbb{R}^D$ that are non-differentiable almost everywhere on their domain \cite{nottale1, falconer, kigami, parv}.\footnote{A fractal curve can be considered as a sequence of piecewise smooth curves $\gamma_i:I=[0,1] \to \mathbb{R}^D$, $i \in \mathbb{N}$, so that for any $t \in I$, the limit of $\frak{F}(t)=\lim_{i\to \infty} \gamma_i(t)$ exists and the Hausdorff-Besicovitch dimension of $\frak F(I)$ exceeds its topological dimension $\text{dim}_T(\frak F(I))=1$.}

\par It is worth noting that by considering the fractal curves as the contour line of self-similar nowhere differentiable maps $\frak F:\mathbb{R}^D \to \mathbb{R}$ the study of fractal curves is transferred to fractal functions on Cartesian spaces. This process is equivalent to transfer from the fractal trajectories in Feynman's path-integral formulation of quantum mechanics \cite{feynman, feynman2} to fractal quantum fields in Feynman's path-integral formulation of quantum field theory.\footnote{See \cite{balankin, calcagni, laskin, kroger, morales} and the references therein for more discussions.} Then, it would be possible to provide a firm understanding of fractal functions which may supply a suitable framework for formulating the established fascinating features of quantum field theories.\footnote{See for example \cite{kroger, shirkov3, wilson2, nottale1}.} Fractal functions as the most basic concepts of the Wilsonian renormalization approach are the core elements of a firm extended geometric framework, usually referred to as \emph{fractal geometry}.

\par Actually, fractal geometry is extensively known as the most common natural feature of quantum mechanics and the fundamental interactions of nature. Fractal structures that show up in most particular observables are actually generated by the dynamics of quantum systems based on the uncertainty principle. In particular, the first appearance of fractal geometry in quantum mechanics was invoked by Feynman to demonstrate the underlying role of self-similarity in the setting of path-integral \cite{feynman}. As mentioned above, the next, but in principle, the most significant emergence of fractal geometry was due to the renormalization group flow in quantum field theory, where the scale (conformal) invariance and critical phenomena in condensed matter physics, and the scale invariance properties in nuclear physics have been thoroughly studied by addressing the problems to the fundamental structure of self-similarity \cite{bogo, wilson1, wilson2}.

\par Thereafter, a vast number of discovered phenomena of quantum physics were properly described by employing the abilities of fractal geometry, such as the critical exponents in condensed matter, the confinement phase in QCD, the Hausdorff dimension of quantum gravity and the Ising model, the critical clusters, and some aspects of spin physics.\footnote{See \cite{kroger} as an interesting research about these topics.} Since fractal geometry plays an extensive role in quantum physics, it may be expected to have a critical contribution in fundamental interactions of nature due to elementary particles. In his fascinating work about fractal geometry and its application in quantum physics \cite{kroger} Kroger asks in this regard: \emph{What is the [background fractal] geometry of propagation for a relativistic quantum particle?} However, since not all structures included in quantum propagations are fractal functions, it seems that to answer this question, we must find a criterion to determine the amount of self-similarity and non-differentiability, the so-called \emph{fractality}, of quantum states or fields.

\par Therefore, the propagation of quantum fields in a renormalizable field theory must be interpreted as the evolution of geometric structures with alternating amounts of fractality in the context of fractal geometry \cite{wilson2, bogo}. On the other hand, there are various elaborated mathematical tools for studying stochastic dynamics (propagation) of geometric structures, wherein the Wiener Brownian process is known as one of the most significant and reliable formulations due.\footnote{See \cite{li} and the references therein for other mathematical procedures.} Therefore, Kroger's question could be perceived as: \emph{What would happen if fractality propagates stochastically via Wiener's formulation of Brownian motion?} This question encourages us to comprehensively explore the Brownian motion of quantum particles in the Hilbert space of quantum states by invoking a consistent norm of fractality that transfers fractal functions to the extremely far distances of the Hilbert space.

\par However, there is no well-defined Wiener Brownian process for fractality in infinite-dimensional Hilbert spaces. Therefore, we should first provide an appropriate criterion for measuring the amount of fractality (i.e. the amount of self-similarity and non-differentiability) of continuous functions and then work out the Wiener stochastic process by addressing fractality. In principle, the more accurate statement of the Kroger's question is:\\

\par \textbf{Question;} \emph{What would be the Wiener path-integral formulation for the Brownian motion of fractality in an infinite-dimensional Hilbert Space?}\\

\noindent Or more precisely:\\

\par \textbf{Question;} \emph{What would be the Wiener path-integral formulation if the contribution of each function in the path-integral is proportional to its amount of fractality?} \\ 

\par In the present research, we will answer the above questions by employing the celebrated Wiener stochastic process \cite{wiener, wiener2} with some modifications based on the amounts of the fractality of continuous functions in the infinite-dimensional Hilbert space of quantum states. At the first step, we will study Hardy's nowhere differentiability condition for the asymptotic behavior of Weierstrass-like functions (section II), and hereby we will work out a specific norm, the so-called \emph{fractal norm}, for measuring the amount of fractality of continuous functions (sections III and IV). The analytical properties of the fractal norm are studied afterward and its physical applications in quantum field theory are discussed (section V). Next, the Wiener measure of the fractal norm, known as the \emph{Wiener fractal measure}, is utilized to figure out the Brownian motion in the infinite-dimensional Hilbert space of quantum states (section VI). Hereby, we will demonstrate that the stochastic propagation of fractality leads to some significant physical results including:\\

\par \textbf{i)} The Wiener fractal measure embraces the Lorentz symmetry at its first local approximation, confirming the Lorentz symmetry as an approximate local symmetry in the Brownian motion of quantum fields (section VII);
\par \textbf{ii)} The local terms of the Wiener fractal measure contain the Klein-Gordon Lagrangian, providing a consistent framework for the Feynmen's path-integral formulation of scalar field theory and the currents understanding of quantum physics due (section VIII);
\par \textbf{iii)} The Wiener fractal measure provides a profound understanding of the machinery of the Wilsonian renormalization group flow (section IX).\\

\par It is worth noting that what makes this research unique and novel is that the above results are essentially based on the foundations of fractal geometry and the Wiener stochastic process without considering unreliable assumptions about Brownian motion in infinite-dimensional Hilbert spaces such as imposing doubtful terms from operator algebra to the standard Wiener's formula and employing some manual modifications of complex analysis in the Wiener stochastic process\footnote{Such as to start with a complex version of the Wiener measure (which is not a probability or finite measure) or to consider the analytic continuation of the mass term within the Lagrangian density.} which all have been extensively studied and accomplished before.\footnote{See for example \cite{bodmann, dau, dau2, dau3, and, nelson, cameron1, cameron2, chaichian, ito1, ito2, gelfand, kac}.} In other words, the present study is distinct from all previous attempts to work out Feynman's path-integral formalism from the Wiener stochastic process, avoiding their four most prevalent, and dubious-sounding, approaches:\\

\par \textbf{1.} \emph{Naively generalizing the standard Wiener measure formula to the intangible process of Brownian motion in an infinite-dimensional Hilbert space.}
\par \textbf{2.} \emph{Carelessly attributing the whole kinetic term of the Lagrangian density to the power of the Gaussian distribution.}
\par \textbf{3.} \emph{Attempting to drag the imaginary number $i$ into the standard formulation of Wiener measure in some incorrect, vague, or enigmatic way.}
\par \textbf{4.} \emph{Generating a custom Lorentz symmetry by manipulating the Gaussian distribution or the role of the invited $i$.}\\

\par Hence, while no deliberate changes, handcrafted modifications, and probably suspicious tricks have been used to the real analytical framework of the Wiener path-integral formula to arrive at the well-known complex formula of Feynman's path-integral, nevertheless, our achievements produce a mathematically well-defined measure that includes some non-local terms but reflects the whole aspects of the current formulation of scalar quantum field theory. In principle, one may be optimistic that the achieved results of this perusal may help us to perceive and obtain more profound intuitions about the foundations of quantum physics and fundamental interactions of nature, including the Yang-Mills fields in the Standard Model and gravity.\footnote{The gravitational fields are studied in \cite{varshovi E-H}, while the Yang-Mills gauge fields are considered in \cite{varshovi}.}


\par
\section{Fractality and Asymptotic Behaviors}
\setcounter{equation} {0}

\par As mentioned above the background fractal geometry of quantum physics is, in fact, referred to \emph{self-similarity} and \emph{nowhere differentiability} of the wave functions or the quantum fields of elementary particles including the Higgs boson. Before constructing a general framework for studying fractals one should revisit the self-similarity and nowhere differentiability conditions first. In principle, these two properties are best understood by considering the Weierstrass function $W(x)=\sum_{k\geq 0}a^k \sin(b^k\pi x/L):I\to \mathbb R$, with $I=[-L,L]$, and $0<a<1$, wherein $b$ is an odd integer \cite{weierstrass}. This function, with no more conditions provides the simplest realization of self-similarity, whereas it was shown by Hardy \cite{hardi} that to gain nowhere differentiability one more condition must be considered too: $ab>1$.\footnote{Weierstrass showed \cite{weierstrass} that the nowhere differentiability would appear when $b/a>1+3\pi/2$, but, subsequently Hardy \cite{hardi} proved this condition could be weakened to $b\geq a^{-1}>1$. However, the case of $b=a^{-1}$ is exotic which leads to almost everywhere smooth functions with discontinuity. Hence, we exclude this case in our arguments, because we are mostly interested in continuous periodic functions. See \cite{johnsen} for a more detailed discussion.} Hence, to obtain both self-similarity and nowhere differentiability the Weierstrass function $W(x)$ must obey a definite power law in its Fourier coefficients: If $n=b^k$, then the $n$-th Fourier coefficient of $W(x)$ is definitely proportional to $1/n^s$ wherein $s=\log_b(1/a)$.\\

\par \textbf{a) Entire Fractals}\\

\par  Actually, self-similarity and nowhere differentiability put a limitation on $s$: $0<s<1$. It is the basic criterion for recognizing the \emph{pure fractality}, i.e. being a summation of a continuous function and a Weierstrass function so that the latter is defined over the entire domain. Roughly speaking, an \emph{entire fractal function} is a continuous map with a power law in its Fourier coefficients $f_n$s as:
\begin{equation} \label {fractal cond}
~~~~~~~~~~~~~~~~~~~~~~~~~~~~~~f_n \propto 1/n^s+\mathcal{O}(1/n^m)~~~~~~~~~~~~~~~~~~~~~(0<s<1)
\end{equation}

\noindent for some definite $b \in \mathbb N$, some integer $m>1$, and for all $n\propto b^k$, $k \geq 1$. We refer to (\ref {fractal cond}) as \emph{Weierstrass ordinary power law} which is abbreviated as OPL.

\par This regular behavior of Fourier coefficients of the Weierstrass function is in principle the immediate consequence of self-similarity. In fact, if a process of $b$-time ($1<b \in \mathbb{N}$) magnification on the $x$-axis results in $a$-time ($0<a<1$) growth along the $y$-axis on the graph of the function $f(x)$, then the Fourier coefficients of $f(x)$ obey a similar condition as (\ref {fractal cond}). In such cases, we say $f(x)$ is an \emph{$(a,b)$-self-similar function}.\footnote{One should note that $(a,b)$-self-similarity will also cause $(a^n,b^n)$-self-similarity. However, in this paper we are mostly interested in the fundamental case which thereby $f(x)$ is $(a,b)$-self-similar if and only if it is not $(a^{1/m},b^{1/m})$-self-similar except for $m=1$.} The pair $(a,b)$ is referred to as the \emph{type of the self-similarity}, and $a$ is called the \emph{scaling factor}. Let us revisit this property with more details: Assume that $f(x):I=[0,2L] \to \mathbb{R}$ is an $(a,b)$-self-similar function and $a_N=\frac{1}{{L}}\int_I f(x)\cos(\frac{{N\pi}}{{L}}x)dx$ be the $N$-th Fourier cosine coefficient. Due to the $(a,b)$-self-similarity the restriction of $f(x)$ to $[k\frac{{2L}}{{b}},(k+1)\frac{{2L}}{{b}}]$, for each $0\leq k <b$, provides the total graph of $f(x)$ multiplied by $0<a<1$ in the $y$-axis and by $1/b$ in the $x$-axis. Hence, if $N=kb$, $k \in \mathbb{N}$, we easily find: $a_N=a a_{k}$, which leads to: $a_n=\frac{{1}}{{n^s}} a_1$, for $n$ being some power of $b \in \mathbb{N}$, whereas $s=\log_b(a^{-1})$.

\par Obviously, the same relation holds for the Fourier sine coefficient $b_N=\frac{{1}}{{L}}\int_I f(x)\sin\left(\frac{{N\pi}}{{L}}x\right)dx$. Thus, according to \cite{hardi} this self-similarity is turned to fractality as one considers the additional condition of nowhere differentiability: $0<s<1$. From now on $s$ is referred to as the \emph{fractality scale}. \\

\par \textbf{b) Widespread Fractals}\\

\noindent One can easily check that the same condition holds for a partial fractal function, i.e. a function $f:I \to \mathbb R$, which is fractal only on a subinterval $\frak I=[x_1,x_2] \subset I=[a,b]$. In fact, partial fractality causes the Fourier coefficients $f_n$ ($=a_n$ or $b_n$) to obey an asymptotic behavior as:\footnote{The proportionality factor could be positive or negative.}
\begin{equation} \label {gw}
~~~~~~~~~~~~~f_n \propto 1/n^s+\mathcal{O}(1/n^m,1/n^{s+m}),~~~~~~~~~~(0<s<1)
\end{equation}
\noindent for $s=\log_b(a^{-1})$, and $n$ a sufficiently large power of $b$, wherein $\mathcal{O}(1/n^{s+m})$ comes from \emph{non-fractal} parts, being of order $1/n^m$ and $1/n^{s+m}$ for $m \geq 1$.

\par The asymptotic behavior of (\ref {gw}) together with condition $0<s<1$, which is assumed to exclude the differentiability on the self-similarity domain, is accordingly referred to as the \emph{Weierstrass peculiar power law} (PPL).\footnote{One should note that, in the PPL, $b>1$ is the least integer fulfilling (\ref {gw}). For example, the Fourier expansion of $f(x)=x:[-1,1] \to \mathbb{R}$, leads to a similar power law $\propto 1/n$, for each $n\in \mathbb{N}$. But, in this case the least integer $b$ is the unity and $s=1$. Therefore, we have no PPL here, hence no fractality involved.} Meanwhile, $f(x)$ may contain several partial fractalities of various types $(a_i,b_i)$. In such cases we see: $f_n\propto \sum_i 1/n^{s_i}+\cdots$, with $0<s_i=\log_{b_i}(a^{-1}_i)<1$. Here, one finds a summation of some PPLs due to (\ref {gw}) for each $s_i$, hence by definition $f(x)$ fulfills the PPL. Moreover, for each fixed $b$, there may be various scaling factors $a_k$s chosen randomly at each Fourier mode, i.e. $n=b^k$ \cite{falconer}. In such cases, we also see that the Fourier coefficients obey the PPL for some specific $0<s<1$.\footnote{Here we have assume $0<a_k<1$ and $a_k b>1$ for each $k$. Thus, the PPL (\ref{gw}) is due to $s=\log_b(a^{-1})$ with scaling factor $a=\lim\text{Sup}_{k\geq 1} \sqrt[k]{a_1 \cdots a_k}$, which by assumption fulfills the condition of: $1/b<a<1$.}

\par The partial fractality and the PPL of (\ref {gw}) could be studied more accurately for \emph{widespread fractal functions}, the term which we use for functions that are self-similar except over a very small subdomain. Indeed, employing Taylor's expansion we can see that a more precise version of PPL holds for the Fourier coefficients. To see this clearly let us assume the \emph{localized Weierstrass function} $W_L(x)=\sum_{k\geq 0} a^k \sin (\frac{b^k} {\frak l} \pi x)$ for $x\in [-\frak l,\frak l]$, and $W_L(x)=0$ for $\frak l <|x| \leq L$, wherein $0<a<1$, $b$ is a positive odd integer and $1<ab$, while $\frac{L- \frak l}L \approx 0$. Hence, we obtain the $n$-th Fourier sine coefficient as;
\begin{equation} \label {ipss}
\begin{gathered}
b_n=\frak l  \sum_{k\geq 0} a^k  \frac{ \sin \left( \pi (b^k-(\frak l/L)n ) \right)}{\pi (b^k - (\frak l/L) n)} - \frak l  \sum_{k \geq 0} a^k \frac{ \sin \left({\pi (b^k + ( \frak l / L) n) }\right)}{ \pi (b^k + (\frak l/L ) n)} \\ \approx \frak l  \sum_{k \geq 0} a^k \frac{ \sin \left({\pi (b^k- ( \frak l / L) n) }\right)}{ \pi (b^k- (\frak l/L ) n)}
 \approx { \frak l} a^K +  F_{K}(L- {\frak l}),~~~~~~~~~
\end{gathered}
\end{equation}

\noindent wherein $F_{K}$ is an analytic function with $F_{K}(0)=0$ and $n=b^K$, for some large $K \in \mathbb N$. In fact;
\begin{equation} \label {F}
F_{K}(\zeta)=\frac{\frak l \sin \left(n\pi \zeta/L \right) }{ \pi } \sum_{0 \leq k \ne K}  \frac{a^k }{ b^k-(\frak l /L) n } \approx \frac{\frak l \sin \left(n\pi \zeta/L \right) }{ \pi } \sum_{0 \leq k \ne K}  \frac{a^k }{ b^k- n }.
\end{equation}

\noindent Actually, $F_K$ splits into two geometric series; an infinite series for $k> K$ and 
a finite summation for $k < K$. Hence, for large enough $n$ (and $K$) we obtain;
\begin{equation} \label {F asymptotic}
\begin{gathered}
F_K(\zeta) \approx \frac{\frak l \sin \left(n\pi \zeta/L \right) }{ \pi } \left( \frac{1}{n} \left( \frac{1-a^K}{1-a} \right) + \frac{a^K}{n} \left( \frac{b}{b-a} \right) \right) ~~~~~~~~\\
=\frac{\frak l \sin \left(n\pi \zeta/L \right) }{ \pi } \left( \frac{1/n-1/n^{1+s}}{1-a}+ \frac{1}{n^{s+1}}\frac{b}{b-a} \right).
\end{gathered}
\end{equation}

\noindent Since the sine function is periodic and we are only interested in asymptotic behaviors of Fourier coefficients, then $\sin \left(n\pi \zeta/L \right) $ must be replaced by some positive constant $0< \alpha<\pi$ times $\zeta/L$, which refers to integers $n_i$, $i \in \mathbb N$, with $n_i\pi\zeta/L=2N\pi + \alpha_i \zeta/L$, for some $N\in \mathbb N$ and $0<\alpha_i \leq \alpha$. Therefore, we see that the PPL included in $F_K$ is of the following form;
\begin{equation} \label {F1}
F_K^{PPL}(\zeta) \approx \frac{\alpha \zeta}{L} \frac{a(1-b)}{(1-a)(b-a)} \frac{1}{n^{1+s}},
\end{equation}

\noindent which leads to;
\begin{equation}
\lim_{K(n) \to \infty} a^{-K} nF^{PPL}_K(\zeta)=-\lambda,
\end{equation}

\noindent for
\begin{equation} \label {lambda-mass}
\lambda = \frac{\alpha \zeta}{L} \frac{a(b-1)}{(1-a)(b-a)} \in \mathbb R_{>0}.
\end{equation}

\noindent Hence, one readily deduces;
\begin{equation} \label {b_n}
\begin{gathered}
b_n \approx \frak l a^K \left(1-\frac{s \mu^2}{n} \right) + \mathcal O(1/n^m)= \frak l \left( \frac{1}{n^s}-s \frac{\mu^2}{n^{s+1}} \right)  + \mathcal O(1/n^m) \\
 \approx \frac{\frak l }{ \left(n^2 + \mu^2 \right)^{s/2}} + \mathcal O(1/n^m),~~~~~~~~~~~~~~~~~~~~~~~~~~~~~~~~~~~~~~~~
\end{gathered}
\end{equation}

\noindent wherein we define the \emph{fractal mass} as; $\mu^2=\lambda /s$. In fact, for widespread fractality condition we are still being encountered with a PPL in the asymptotic behavior;\footnote{The proportionality factor could be positive, negative, or complex.}
\begin{equation} \label {gw1}
\begin{split}
~~~~~~~~~~~~~~~~~~~f_n \propto 1/(n^2+\mu^2)^{s/2}+\mathcal{O}(1/n^m).~~~~~~~~~~~~~~~(0<s<1)
\end{split}
\end{equation}

\noindent We may refer to (\ref {gw1}) as the \emph{Weierstrass massive power law} (MPL). In fact, discovering some MPL behavior among the Fourier coefficients is a fiercely complicated process. However, although the PPL will produce the MPL, as we will demonstrate in the next section the MPL condition leads to a more accurate and distinct criterion for fractal functions. Moreover, since the fractal mass $\mu$ is proportional to $\zeta=L- \frak l$, the asymptotic behavior of widespread fractals in (\ref {gw1}) naturally leads to that of entire fractals we worked out in (\ref {fractal cond}) as $\zeta \to 0$.


\par
\section{Fractal Norm and Fractal Functions}
\setcounter{equation} {0}

\par However, one can work out a criterion to discover and figure out the OPL and MPL behaviors for entire and widespread fractal functions. This could be done by putting an appropriate measure on the phase space via the Fourier coefficients.\\

\par \textbf{a) Entire Fractals}\\ 

\par Let us first concentrate on the Weierstrass function. A direct calculation shows that 
by assuming $n$ as a continuous parameter we obtain from (\ref{fractal cond}) the asymptotic inequality\footnote{The coefficient $1/2$ is considered for upcoming formulations, otherwise, we can absorb it in $\kappa$.}
\begin{equation} \label {criteria1'}
\frac{{d|f_n|}}{{dn}}+ \frac{{|f_n|}}{{n}}>|f_n|~e^{-\kappa n^2 |f_n|^2/2},
\end{equation}

\noindent for each function $f(x)$ with entire fractality (i.e. OPL), wherein $\kappa>0$ is an arbitrary constant. Note that here $f(x)$ may be considered as a complex valued function and $f_n$ could be assumed as the complex Fourier coefficient obtained for the Fourier mode $e^{in\pi x/L}$. Hence, $f_n$ is generally a complex number. To consider any kind of self-similarity with assuming the extreme case of non-differentiability, i.e. $s\to 1$, we conclude the inequality 
\begin{equation} \label {criteria2}
\frac{{d|f_n|}}{{dn}}+ \frac{{|f_n|}}{{n}}>\frac{{\ell}}{{n}} ~e^{-\kappa n^2 |f_n|^2/2},
\end{equation}

\noindent which must be hold asymptotically for any fixed constants $\ell,\kappa>0$.\footnote{This inequality is easily inferred by comparing the heat kernel function and the powers of $1/n$.} That is, for each positive $\ell$ and $\kappa$ there exists a large enough $N \in \mathbb{N}$, so that for any $n\geq N$ the inequality (\ref {criteria2}) holds for each function with entire fractality. In principle, (\ref{criteria2}) could be assumed as the main criterion for identifying the entire fractality. In other words, if the equation
\begin{equation} \label {criteria3'}
\frac{{d|f_n|}}{{dn}}+ \frac{{|f_n|}}{{n}}|\leq\frac{{\ell}}{{n}} ~e^{-\kappa n^2 |f_n|^2/2}
\end{equation}

\noindent is imposed to the Fourier coefficients $f_n$s for some fixed $\kappa$ and $\ell$, the resulting function will not admit the entire fractality.

\par Let $\kappa$ be fixed. Then, to obtain a bound for avoiding the entire fractality, we should first solve the differential equation\footnote{See \cite{str} for other formulations of fractals with differential equations.}
\begin{equation} \label {criteria3''}
\frac{{d|f_n|}}{{dn}}+ \frac{{|f_n|}}{{n}}=\frac{{\ell}}{{n}} ~e^{-\kappa n^2 |f_n|^2/2}.
\end{equation}

\noindent The solution of (\ref {criteria3'}) gives $|f_n|$ as a function of $\ell$ and $n$, i.e. $|f_n|=|f_n|(n, \ell)$. When we fix $|f_n|$, then according to the implicit function theorem $\ell$ is given as a function of $n$ and $|f_n|$. This function is given by the following integral formula:
\begin{equation} \label {criteria4}
\ell=\ell_\kappa(f_n)=\int_0^{|f_n|}~e^{\kappa n^2 x^2/2}~dx,
\end{equation}

\noindent where $f_n$ is also allowed to admit complex values.

\par In principle, (\ref {criteria4}) could be easily confirmed by employing the implicit function theorem. To see this let $f_n$ be real and positive, otherwise replace $f_n$ by $-f_n$. Now, according to (\ref {criteria4}) we obtain:
\begin{equation} \label {moadele}
\frac{\partial \ell}{\partial n}+ \frac{\ell}n =\frac {f_n}{n} e^{\kappa n^2 f^2_n/2},~~~~~~~~~~\text{and;}~~~~~~~~~~\frac{ \partial \ell}{\partial f_n}=e^{\kappa n^2 f^2_n/2}.
\end{equation}

\noindent Hence, the differential equation of $f_n$, i.e. (\ref {criteria3''}), could be simply derived with considering $\partial f_n /\partial n=-(\partial \ell/\partial n)/(\partial \ell/\partial f_n)$, which in its own right is the immediate consequence of fixing $f_n$ or equivalently $df_n (n, \ell(n))/dn=0$. We refer to $\ell_\kappa(f_n)$ in (\ref {criteria4}) as the \emph{fractal $\kappa$-norm of $f_n$}, while generally we may drop the prefix $\kappa$, and use simply the \emph{fractal norm} for convenience.

\par As mentioned above the fractal norm of $f_n$ has critical importance in disclosing the existence of the entire fractality inside a continuous function. In fact, to exclude the entire fractality we should have $f_n \to 0$ at least with the rate induced from (\ref {criteria4}) for some fixed $\kappa$ and $\ell$. All in all, we have the following theorem:\\

\textbf{Theorem 1;} \emph{Assume that $I \subset \mathbb{R}$ is a compact interval and $f(x):I \to \mathbb{C}$ is a continuous function with Fourier coefficients $f_n$. Then, $f(x)$ is not an entire fractal function (i.e. dismisses OPL) if and only if for some fixed $\kappa>0$ and $\ell_0>0$ we have: $\ell_\kappa(f_n)\leq \ell_0$, for all $n$. In other words, $f(x)$ is not an entire fractal function if and only if there exists some $\kappa>0$ and $\ell_0>0$ so that the inequality}
$$\int_0^{|f_n|}~e^{\kappa n^2 x^2/2}~dx\leq \ell_0.$$
\noindent \emph{holds for any Fourier coefficient $f_n$. Consequently, $f(x)$ is an entire fractal function if for any $\kappa>0$ and $\ell_0>0$ there exists a Fourier coefficient $f_n$ so that:}
$$\int_0^{|f_n|}~e^{\kappa n^2 x^2/2}~dx > \ell_0.$$\\

\par As a corollary of \textbf{Theorem 1} the function $f(x):I \to \mathbb{C}$ does not comprise any entire fractality behavior if and only if the fractal norm of its Fourier coefficients would produce a bounded set.\\ 

\par \textbf{b) Widespread Fractals}\\

\par For a widespread fractal with fixed $a$, $b$, and $\frak l$, we can easily see from (\ref {gw1}) that;
\begin{equation} \label {local 1}
\frac{{d|f_n|}}{{dn}}+ \frac{{n}}{{n^2+\mu^2}} |f_n| \geq  |f_n| ~e^{-\kappa (n^2+\mu^2)|f_n|^2/2},
\end{equation}

\noindent Consequently, for each widespread fractal function fulfilling the inequality $0<s<1$, we find the following asymptotic equation:
\begin{equation} \label {++}
\frac{{d|f_n|}}{{dn}}+ \frac{{n}}{{n^2+\mu^2}} |f_n|  \geq \frac{{n \ell}}{{n^2+\mu^2}} ~e^{-\kappa (n^2+\mu^2) |f_n|^2/2},
\end{equation}

\noindent where $\ell>0$ is a constant. Therefore, with the similar discussion we had for entire fractals, to find a measure on the phase space that excludes widespread fractal functions we have to solve the following differential equation:
\begin{equation} \label {local 2}
\frac{{d|f_n|}}{{dn}}+ \frac{{n}}{{n^2+\mu^2}} |f_n| =\frac{{n \ell}}{{n^2+\mu^2}} ~e^{-\kappa (n^2+ \mu^2) |f_n|^2/2}.
\end{equation}

\noindent The solution of (\ref {local 2}) works out $|f_n|$ as a function of $\ell$ and $n$, i.e. $|f_n|=|f_n|(n, \ell)$. Upon a similar reasoning to the discussion we had for the entire fractals $\ell$ is given as a function of $f_n$ and $n$:
\begin{equation} \label {local 3}
\ell=\ell_{(\kappa,\mu)}(f_n)=\int_0^{|f_n|}~e^{\kappa (n^2+\mu^2) x^2/2}~dx,
\end{equation}

\noindent where $f_n$ is allowed to admit complex values. The fractal norm $\ell_{(\kappa,\mu)}$ is conventionally called the \emph{fractal $(\kappa,\mu)$-norm of $f_n$}, while we may drop the prefix $(\kappa,\mu)$ for simplicity and use the \emph{massive fractal norm} as the convenient terminology. In particular, this generalized norm provides a criterion for recognizing non-fractal functions. The next theorem could be considered as the immediate generalization of \textbf{Theorem 1} for widespread fractals.\\

\par \textbf{Theorem 2;} \emph{Assume that $I \subset \mathbb{R}$ is a compact interval and $f(x):I \to \mathbb{C}$ is a continuous widespread function with Fourier coefficients $f_n$. Then, $f(x)$ has no widespread fractality if and only if there are some fixed $\kappa>0$, $\mu>0$, and $\ell_{0}>0$, so that we have: $\ell_{(\kappa,\mu)} (f_n)\leq \ell_{0}$, for all $n$. In other words, $f(x)$ has no widespread fractality if and only if there exists some $\kappa>0$, $\mu>0$, and $\ell_{0}>0$, so that the inequality}
$$ \int_0^{|f_n|}~e^{\kappa (n^2+\mu^2) x^2/2}~dx\leq \ell_{0}.$$
\noindent \emph{holds for any Fourier coefficient $f_n$. Consequently, $f(x)$ admits widespread fractality if and only if for any $\kappa>0$, $\mu>0$, and $\ell_0>0$, there exists a Fourier coefficient $f_n$ which fulfills the following inequality:}
$$ \int_0^{|f_n|}~e^{\kappa (n^2+\mu^2) x^2/2}~dx > \ell_{0}.$$


\par
\section{Fractal Norm in Higher Dimensions}
\setcounter{equation} {0}

\par The same ideas could be similarly assumed for functions with $D$-dimensional domains ($D>1$) by multiplying $D$ copies of the Weierstrass function for each variable $x^i$, $1\leq i \leq D$. For instance, in $D=2$ dimensions, we easily construct a self-similar nowhere differentiable function as: $W^{(2)}(x,y)=\sum_{m,n}a^ma'^n\cos(b^m\pi x) \cos(b'^n \pi y)$, with $0<a,a'<1$, and odd integers $1<b,b' \in \mathbb{N}$, fulfilling: $ab,a'b'>1$. In such cases we also see a OPL as: $a_{n,n'} \propto n^{-s}n'^{-s'}$, for $a_{n,n'}$ the Fourier cosine coefficient for $2$-dimensional Fourier mode $(n,n')$, while $n$ (resp. $n'$) is some power of $b$ (resp. $b'$) and $0<s=\log_b(a^{-1})<1$ (resp. $0<s'=\log_{b'}(a'^{-1})<1$). Generally, one may simply define $W^{(D)}(x^1,\cdots,x^D)$ to be produced by multiplying $D$ copies of the Weierstrass function, each of which is an $(a_i,b_i)$-self-similar function, with $b_i \in \mathbb N$ an odd integer, and $0<a_i<1$, the $i$-th scaling factor, while $a_ib_i>1$ for each $1\leq i \leq D$. Substantially, if for $D$-plets $u=(u_1,\cdots,u_D)$ and $v=(v_1,\cdots,v_D)$ one simply sets: $u^v=u_1^{v_1}\cdots u_D^{v_D}$, then $W^{(D)}(x^1,\cdots,x^D)$ admits the OPL in its Fourier coefficients: $f_n \propto1/n^s$, for $n=(n_1,\cdots,n_D)$ be a $D$-plet of integers, and $s$ the $D$-plet power: $s=(s_1,\cdots,s_D)$, with $s_i=\log_{b_i}(a_i^{-1})$. 

\par Also, we can go farther and define PPL accordingly for $D$ dimensions. Thus, if $f(x^1,\cdots,x^D)$ partly admits such self-similarities, together with non-differentiability, then $f_n \propto 1/n^s+\mathcal O(1/n^{s+m},1/n^m)$, $m\geq 1$, for the general Fourier coefficient $f_n$ and for $n_i$ some power of $b_i$, $1\leq i \leq D$. It is the PPL for functions with $D$-dimensional domains.\footnote{Here, we also assume that both self-similarity and nowhere differentiability hold for each of the $D$ directions.} Therefore, if $f$ possess the widespread fractality in some of its $D$ variables, then it fulfills MPL asymptotically in its Fourier coefficients; i.e. $f_n \propto \prod_{i=1}^D 1/(n_i^2+\mu_i^2)^{s_i/2}+\mathcal O(1/n^{m+s_i},1/n^m)$, $m\geq 1$. Specifically, for \emph{symmetric fractals} which admit the same type of self-similarity in each of the variables one discovers a beautiful formula for the leading term in the asymptotic behavior of the Fourier coefficients. To see this let $f$ be a widespread (or entire) symmetric fractal with the fractal mass $\mu_1=\cdots =\mu_D=\mu$ and the fractality scale $s_1=\cdots =s_D=s$. Hence, at the leading term we find: $f_n \propto \prod_{i=1}^D \frac{1}{(n_i^2+\mu^2)^{s/2}}$. This proportionality gives rise to a critical formula which helps us to work out the corresponding differential equations for Fourier coefficients $f_n$. Set $1\leq i \leq D$ and fix $n_j$s for all $j \ne i$. Essentially, here we concern about the asymptotic behavior in the $i$-th direction. Since $0<s<1$ then we asymptotically have;
\begin{equation} \label {sym-0}
\frac{n_i^2+\mu^2}{n^2+\mu^2}>s,
\end{equation}

\noindent wherein $n^2=\sum_{i=1}^D n_i^2$. Therefore, by considering (\ref {sym-0}) and employing the argumentation of the last section we have the following asymptotic behavior of MPL:
\begin{equation} \label {sym-1}
\frac{{\partial | f_n|}}{{\partial n_i}}+ \frac{{n_i}}{{n^2+\mu^2}}|f_n| \geq \frac{{n_i \ell}}{{n^2+\mu^2}} ~e^{-\kappa (n^2 + \mu^2) |f_n|^2/2},
\end{equation}

\noindent wherein $\kappa>0$ is fixed. Hence, we should consider the following system of PDEs to find a criterion to avoid the fractality in $D$-dimensions:
\begin{equation} \label {criteria6}
~~~~~~~~~~~~~~~~~~~~\frac{{\partial |f_n|}}{{\partial n_i}}+ \frac{{n_i}}{{n^2+\mu^2}}|f_n|=\frac{{n_i\ell}}{{n^2+\mu^2}} ~e^{-\kappa (n^2+\mu^2) |f_n|^2/2}.~~~~~~~~~~~~~(1\leq i \leq D)
\end{equation}

\noindent Consequently, (\ref {local 3}) is also obtained in $D$-dimensions for the massive fractal norm $\ell_{(\kappa,\mu)}$ as;
\begin{equation} \label {D dimension 1}
\ell = \ell_{(\kappa,\mu)}(f_n)=\int_0^{|f_n|}~e^{\kappa (n^2+\mu^2) x^2/2}~dx=\int_0^{|f_n|}~e^{\kappa (n_1^2+\cdots +n_D^2+\mu^2) x^2/2}~dx.
\end{equation}

\noindent Obviously for the entire fractals we should set $\mu=0$ and replace $\ell_{(\kappa,\mu)}$ with $\ell_\kappa=\ell_{(\kappa,0)}$. In principle, we simply read;\\

\par \textbf{Corollary 1;} \emph{Assume that }$V \subset \mathbb{R}^D$ \emph{is a compact $D$-dimensional cube and} $f(x^1, \cdots ,x^D): V \to \mathbb{C}$ \emph{is a continuous function with Fourier coefficients} $f_n$\emph{. Then, }$f ( x^1, \cdots , x^D )$ \emph{has no widespread (resp. entire) fractality if and only if there exists some} $\ell_0>0$, $\kappa>0$ \emph{and} $\mu > 0$\emph{, so that} $\ell_{(\kappa , \mu)}(f_n) \leq \ell_0$ \emph{(resp. }$\ell_{\kappa}(f_n)\leq \ell_0$\emph{) for each }$f_n$.\\

\par To enjoy the simplicity of the Fourier analysis on cubes, we are mostly interested in functions with cubic domains. Even, if the domain of the function $f(x^1,\cdots,x^D)$ is a bounded $D$-dimensional region, say $U\subset \mathbb{R}^D$, we prefer to restrict it to some large enough cube $V$ which is included in $U$ as a proper subset. Then the continuous function $f:U\to \mathbb{C}$ is said to be a fractal function if there exists a cube $V\subset U$, on which $f$ is self-similar and nowhere differentiable except on a subset with a very small Lebesgue measure. However, a more strict definition of fractal function is:\\

\par \textbf{Definition 1;} \emph{Assume that $U\subset \mathbb{R}^D$ is a bounded $D$-dimensional set and let $f(x^1,\cdots,x^D):U \to \mathbb{C}$ be a continuous function. Equivalently, $f$ is referred to as a fractal function if there exists a cube $V\subset U$ so that $\{\ell_{(\kappa,\mu)}(f_n)\}$ is an unbounded set, wherein $f_n$ is the $n$-th Fourier coefficient of $f$ restricted to $V$.}


\par
\section{Analysis and Applications of Fractal Norm}
\setcounter{equation} {0}

\par Let $S=[-L,L]^D\subset \mathbb R^D$ be a compact $D$-dimensional cube\footnote{For simplicity, we have assumed that $S$ has the same length in all coordinates. However, the generalization of this assumption is immediate.} and denote the normalized Laplacian eigenfunctions with $\{ \phi_n \}_{n \in \mathbb Z^D}$. We can simply assume that in each $n=(n_1,\cdots,n_D)\in \mathbb{Z}^D$, a non-negative integer, say $n_i$ ($1\leq i \leq D$), corresponds to the cosine function $\cos\left(\frac{n_i \pi}{L}x^i\right)$, and a negative one, say $n_j$ ($1\leq j \leq D$), is related to the sine function $\sin\left(\frac{|n_j| \pi}{L}x^j\right)$. Set $\mathbf C \subset C^0(S)$ to be the set of all continuous functions that satisfy the following condition
\begin{equation} \label {moad-1}
f(x)=\sum_{n \in \mathbb Z^D} f_n \phi_n(x)
\end{equation}

\noindent for almost every $x \in S$. According to the Fourier's theorem\footnote{See \cite{Fou}.} $\mathbf C$ contains $C^1(S)$, whereas there is no guarantee for $f \notin C^1(S)$ to satisfy (\ref {moad-1}). But, however, the Weierstrass function belongs to $\mathbf C$ while it is out of $C^1(S)$. Hence, $\mathbf C$ contains some continuous functions with peculiar properties such as fractality. Employing \textbf{Definition 1} we can define a useful metric, the so-called \emph{fractal metric}, on $\mathbf C$ and turn it to a metric space with some interesting topological properties:\\

\par \textbf{Definition 2;} \emph{Assume that $f,g \in \mathbf C$, and fix $\mu \geq 0$, $\kappa >0$. Then, the fractal metric $d_{\mu,\kappa}(f,g)$ is defined as;}
\begin{equation} \label {moad-2}
d_{\kappa,\mu}(f,g)=\text{Sup}_{n \in \mathbb Z^D}\Big|  \int_{g_n}^{f_n} e^{\kappa ( n^2 + \mu^2)x^2/2} dx     \Big|,
\end{equation}

\noindent \emph{wherein the supremum is defined over all Fourier modes $n \in \mathbb Z^D$.}\\

\par It can be easily seen that according to \textbf{Definition 1} and \textbf{Definition 2} the following holds for each $f,g,h \in \mathbf C$;\\
\par \textbf{i)} $d_{\kappa,\mu}(f,g) \geq 0$ (\emph{non-negativity});
\par \textbf{ii)} $d_{\kappa,\mu}(f,g)=0$ if and only if $f=g$ (\emph{identity of indiscernibles});
\par \textbf{iii)} $d_{\kappa,\mu}(f,g)=d_{\kappa,\mu}(g,f)$ (\emph{symmetry});
\par \textbf{iv)} $d_{\kappa,\mu}(f,g)\leq d_{\kappa,\mu}(f,h)+d_{\kappa,\mu}(h,g)$ (\emph{triangle inequality}).\\

\noindent The above properties ensure that $d_{\kappa,\mu}$ is a well-defined metric, hence $(\mathbf C,d_{\kappa,\mu})$ is a metric space. On the other hand, according to \textbf{Definition 2} the convergence in fractal metric $d_{\kappa,\mu}$ automatically leads to the point-wise convergence in the phase space. That is, if $\{ f^{(i)}\}_{i=1}^\infty$ is a Cauchy sequense in $(\mathbf C,d_{\kappa,\mu})$, then for each $n \in \mathbb Z^D$ the Fourier coefiicients $\{ f_n^{(i)}\}_{i=1}^\infty$ provide a convergent sequence. Let $f_n^{(i)} \to f_n$ and set $f(x)=\sum_{n\in \mathbb Z^D} f_n \phi_n(x)$. The convergence of $\{ f^{(i)}\}_{i=1}^\infty$ ensures that $f(x)$ exists for almost all $x \in S$. Hence, upon Egoroff's theorem,\footnote{See \cite{Princeton, Springer}.} $f$ belongs to $\mathbf C$. Indeed, the limit of any Cauchy sequence in $(\mathbf C, d_{\kappa,\mu})$ converges to an almost everywhere continuous function on $S$ which satisfies (\ref {moad-1}). Thus, $\mathbf C$ is topologically a complete set and contains all its limit points. This beneficial analytic property will be used to work out our main formula for the Wiener measure of $\mathbf C$.

\par In addition, according to \textbf{Definition 2} we have two more interesting properties for fractal functions in $\mathbf C$:\\
\par \textbf{v)} $d_{\kappa,\mu}(f,g)<\infty$ if $f$ and $g$ are both non-fractal;
\par \textbf{vi)} $d_{\kappa,\mu}(f,g)=\infty$ if $f$ is fractal and $g$ is non-fractal.\\

\par Hence, upon properties \textbf{(v)} and \textbf{(vi)} and according to the metric topology of $(\mathbf C, d_{\kappa,\mu})$, the fractal functions are located in the extremely far distant region of the set $\mathbf C$. This distance is defined according to the fractal norm (\ref {D dimension 1}). In fact, the fractal metric measures the amount of fractality of continuous functions. Actually, we have a quantative definition of \emph{fractality}:\\

\par \textbf{Definition 3;} \emph{Let $f_0 =0 \in \mathbf C$ be the zero function. Then, the $(\kappa,\mu)$-fractality of $f \in \mathbf C$ or simply the fractality of $f \in \mathbf C$ is defined as $d_{\kappa,\mu}(f,f_0)\in [0,+\infty]$.}\\

\noindent This, actually, means that by considering the fractal norm $\ell_{(\kappa,\mu)}$ in studying some asymptotic aspects via the limiting process of $\ell_{(\kappa,\mu)} \to \infty$, we will be able to do two following actions:

\par \textbf{a)} To restrict the domain of calculations to $\mathbf C$;

\par \textbf{b)} To include the role of fractality in our perusal. 

\noindent We should emphasize that these two processes \textbf{(a)} and \textbf{(b)} are necessary for studying the theory of quantum fields. In fact, \textbf{(a)} ensures that the quantum fields have proper Fourier expansions satisfying (\ref {moad-1}), which is the most fundamental hypothesis of the second quantization. However, \textbf{(b)} is the main motivation for renormalization group flow in quantum field theories. Actually, if we define the Wiener path-integral measure by employing the fractal norm $\ell_{(\kappa,\mu)}$ then the path-integral measure would be intrinsically compatible with the Wilsonian renormalization process. Moreover, as we will see in the following, this will, in fact, turn the Wiener path-integral measure into the well-known complex measure of Feynman's path-integral formulation of scalar quantum fields in a mathematically smooth way.

\par Without considering the fractal metric, the Wiener path-integral domain will contain functions that neither fulfill the condition (\ref {moad-1}) nor are even continuous. Consequently, it seems that to have a well-defined framework for the path-integral formulation of quantum field theories one is obliged to consider the induced measure of the fractal norm on the phase space of functions of $\mathbf C$ as the domain of the Wiener path-integral. This is, in fact, the immediate consequence of the metric topology of $(\mathbf C, d_{\kappa,\mu})$. Hence, by employing the fractal norm and its induced measure/topology the domain of the Wiener path-integral coincides with that of Feynman's formulation.

\par In other words, the fractal norm lets us regard the fractality whenever we are concerned about their contributions. In fact, as we will see in the next section, employing the fractal norm and its induced Lebesgue measure on the phase space of the Foureir modes causes the fractality to be included in the calculations via a smooth consistent process due to the exponential of the Laplacian eigenvalues. Thus, the fractal norm (\ref {D dimension 1}) would provide an appropriate machinery to consider the involved fractality in studying both statistical and stochastic properties of the time evolution of continuous functions in the Hilbert space of quantum states. Therefore, in order to include the fractal structures within the Wiener stochastic process of continuous functions one has to write down the Wiener path-integral formula in terms of the fractal norm and its induced measure on the phase space. In the upcoming sections, we will see that this approach actually leads us from Wiener's path-integral to Feynman's path-integral automatically.

\par After figuring out the fractal norm, its induced metric, and its emergent topological features, it would be fine to provide here a descriptive discussion to explain the mathematical formulations and physical implications of what has been achieved up to now and what is to be extracted later in the rest part of the article.

\par Upon the above arguments, here we prefer to provide a comprehensive and detailed explanation of the terminological meaning of the \emph{fractal norm} and \emph{fractal metric}:\\

\par \textbf{Remark 1) The Meaning of Fractal Norm and Fractal Metric;} \emph{We should note that the "fractal norm" is not, in fact, a norm for fractals, but it is the criterion that effectively measures the amount of fractality of continuous functions via some analytical features. Thus, the fractal metric is actually a measurement for continuous functions which shows how far a continuous function has come in self-similarity and non-differentiability to become a fractal function.}\\

\par Thus, according to \textbf{Remark 1} any continuous function in $\mathbf C$ possesses fractality as long as its fractal metric testifies. Hence, any element of $\mathbf C$ could be effectively considered a continuous function with more or less fractality. It is exactly the main idea of the renormalization group process. This viewpoint will be addressed for formulating the Wiener process in the infinite-dimensional Hilbert space of quantum states. We may refer to it as the \emph{Wiener process of fractaity}. Thus, we have:\\

\par \textbf{Remark 2) The Meaning of Wiener Process of Fractality;} \emph{One should note that in the expression of "Wiener process of fractality" all propagating elements in $\mathbf C$ are considered from the viewpoint of fractality that is evaluated by means of the fractal norm. In other words, the expression addresses the Wiener formulation of the stochastic process of the continuous functions in $\mathbf C$ when the location of each function is identified by its progress in fractality.}\\

\par Actually, as we will establish in the upcoming sections of this research, the fractal norm together with its developed structures, such as its induced measure on $\mathbf C$, is actually the most appropriate mechanism by which the fractal aspects, as the most transcendent analytic property and also the most ultimate asymptotic behavior of continuous functions, are entered smoothly in our main calculations. Hence, as explained in \textbf{Remark 2}, the Wiener process of fractality contains the whole contributions of elements of $\mathbf C$ upon their features of self-similarity and non-differentiability. Such functions, from the viewpoint of quantum physics, are indeed compared with quantum fields involved in the Feynman's path-integral formulation. Therefore, we claim that:\\

\par \textbf{Remark 3) The First Purpose of Considering the Fractality;} \emph{By considering the fractality within the Wiener stochastic process we basically aim to work out a Wiener path-integral formulation for quantum fields so that the path-integral measure intrinsically contains fractality and is best compatible with the Wilsonian renormalization process.}\\ 

\par One should note that accomplishing the process of \textbf{Remark 3} will produce some non-local terms within the genuine form of Feynman's path-integral measure of quantum field theories. In other words, considering the whole effects of fractality within the path-integral measure gives rise to appearing non-local aspects in the Gaussian term of the Wiener measure. Although it may be possible to transfer the contribution of these non-local features to some approximating local terms on flat space-time, their important role in controlling the infinities of the theory could not be easily ignored when one is studying the gravitational effects in quantum physics. Thus:\\

\par \textbf{Remark 4) The Ultimate Goal of Considering the Fractality;} \emph{Although considering the fractality would produce non-local terms within the path-integral formulation of quantum fields and their interactions, it will lead to automatically finite, hence predictable theories via some mathematically consistent approach. Consequently, we are interested in the question of whether this method can provide a way to produce a well-defined theory of quantum gravity by means of an appropriate path-integral formulation.}
 

\par
\section{Wiener Fractal Measure and Path-Integrals}
\setcounter{equation} {0}

\par It is well-known that the Brownian motion of a particle in $D$-dimensional space is given by means of the celebrated Wiener probability measure. More precisely, if the particle is at $x_0 \in \mathbb R^D$ at $t_0$, then the probability of the particle being found in $U_i \subset \mathbb R^D$ at $t_i$, $1\leq i \leq N$, with the time ordering of $t_0<t_1<\cdots <t_N$ is given as \cite{wiener, wiener2};\footnote{See also \cite{folland, durret} as two nice presentations of the issue. Moreover, grasping the elegant viewpoint of \cite{gol} is beneficial.}
\begin{equation}
\begin{gathered}
P(U_1,\cdots, U_N;t_1,\cdots,t_N)
=\left( \frac{1}{2\pi \tau_N}\right)^{D/2} \cdots \left( \frac{1}{2\pi \tau_1} \right)^{D/2} \\ \times \int_{U_N}\cdots \int_{U_1} e^{-|x_N-x_{N-1}|^2/2\tau_N}\cdots e^{-|x_2-x_1|^2/2\tau_2} e^{-|x_1-x_0|^2/2\tau_1}~d^Dx_N \cdots d^Dx_1,
\end{gathered}
\end{equation}

\noindent wherein $\tau_i=t_i-t_{i-1}$, $1\leq i\leq N$. Also, we used $x_i=(x_i^1,\cdots,x_i^D)$ and $d^Dx_i$ respectively for the coordinate system and the Lebesgue measure of $\mathbb R^D$ at $t_i$ (i.e. $x_i=x(t_i)$), and employed the Euclidean metric $|x|^2=(x^1)^2+\cdots+(x^D)^2$ for $\mathbb R^D$. In other words, the Wiener measure for the time slicing $t_0<t_1<\cdots<t_N$ is;
\begin{equation} \label {wiener1'}
dW(t_N,\cdots,t_1)=\left( \frac{1}{2\pi \tau_N} \right)^{D/2} ~ e^{-|x_N-x_{N-1}|^2/2\tau_N} ~d^Dx_N \times \cdots \times \left( \frac{1}{2\pi \tau_1} \right)^{D/2} ~ e^{-|x_1-x_0|^2/2\tau_1} d^Dx_1.
\end{equation} 

\noindent One may be interested in considering the symmetric form of the Wiener measure as:
\begin{equation} \label {wiener1}
\begin{gathered}
dW(t_N,\cdots,t_1)=\left( 4\pi T \right)^{D/2}~ \exp \left( |x_F-x_I|^2/4T \right)\\
\times \left( \frac{1}{2\pi \tau_{N+1}} \right)^{D/2}~ e^{-|x_F-x_{N}|^2/2\tau_{N+1}} ~ d^Dx_N ~ \left( \frac{1}{2\pi \tau_N} \right)^{D/2} ~ e^{-|x_N-x_{N-1}|^2/2\tau_N} ~ \times \cdots \\
 \times \cdots \times \left(  \frac{1}{ 2\pi \tau_1 } \right)^{D/2} ~e^{-|x_2-x_1|^2/2\tau_2} ~ d^Dx_1 ~ \left( \frac{1}{2\pi \tau_1}\right)^{D/2}~e^{-|x_1-x_I|^2/2\tau_1},
\end{gathered}
\end{equation} 

\noindent which evaluates the corresponding amount of probability of a particle initiating its Brownian motion from $x_I$ at $t_0=-T$, and terminating its random journey at $t_{N+1}=T$ with reaching $x_F$. Obviously, we easily see that:
\begin{equation}
\int_{\mathbb R^D} \cdots \int_{\mathbb R^D} dW(t_N,\cdots,t_1)=1,
\end{equation}

\noindent which ensures that $dW$ is a probability measure.

\par In the following, we only consider the symmetrized form of the Wiener measure (\ref {wiener1}) to work out the physical interpretations due to quantum field theory. Hence, if the particle is subject to some external force given in terms of some measurable functions $e^{V_i(x)}$ for potentials\footnote{To guarantee the integrability the potential functions $V_i(x)$s must be bounded from above.} $V_i(x)$, $1\leq i \leq N$, within the partition of time $-T<t_1<\cdots <t_N<T$, then the influence of $V_i$s on the particle being located in $U_i \subset \mathbb R^D$ at $t_i$, $1\leq i \leq N$, is studied by calculating the following integral:
\begin{equation} \label {weiner2}
P(U_1,\cdots, U_N;t_1,\cdots,t_N) = \int_{U_N}\cdots \int_{U_1} e^{-V_N(x_N)} \cdots e^{-V_1(x_1)} dW(t_N,\cdots,t_1).
\end{equation}

\noindent However, the probabilistic interpretation of the above Wiener integral entirely depends on the theory. Also, the term of particle which we used above addresses essentially an element of the target space $\mathbb R^D$. Hence, if $\mathbb R^D$ is replaced by the space of some special Fourier modes satisfying some specific conditions, the above Brownian motion interpretation of the Wiener integral must be translated to the concepts correlated to the space of the corresponding functions.

\par It is worth to note that the sufficient condition for potentials $V_i$ in (\ref {weiner2}) to lead to some finite values for $P(U_1,\cdots, U_N;t_1,\cdots,t_N)$ is;
\begin{equation} \label {con}
\int_{\mathbb R^D} e^{-a|x-x_0|^2+V_i(x)}d^Dx<\infty
\end{equation}

\noindent for any $a>0$ and $x_0\in \mathbb R$, which is certainly fulfilled by potentials that are bounded from above. We may refer to (\ref {con}) as the \emph{Wiener convergence condition}. Moreover, assuming the condition (\ref {con}) the measurable function $F(x_N(t_N),\cdots,x_1(t_1))=e^{V_N(x_N(t_N))} \cdots e^{V_1(x_1(t_1))}$, and each of the potentials $V_i(x)$ are commonly called \emph{Wiener convergent functions}. 

\par Now, let $\mathcal N \in \mathbb N$, set $S=[-L,L]^D \subset \mathbb R^D$, and define $\mathbf C_{\mathcal N}$ to be the space of continuous functions $f:S\to \mathbb R$ in $\mathbf C$ with the Fourier coefficients $f_n$ which vanish except for Fourier modes $n= (n_1,\cdots,n_D )$ satisfying $-\mathcal N \leq n_i \leq \mathcal N$ for all $1\leq i \leq D$. Here, for simplicity and without loss of generality one can assume that each $f_n$ is a real number. For instance, one may suppose that a negative (resp. non-negative) component of $n=(n_1,\cdots,n_D)$ corresponds to a sine (resp. cosine) Fourier coefficient in that dimension. Similarly, for the Fourier complex coefficients calculated for the Fourier basis $\{e^{i\pi n.x/L}\}_{n\in \mathbb Z^D}$, the Fourier coefficients $f_n$ and $f_{-n}$ provide two independent real variables as $\frac{f_n+f_{-n}}{2}$, and $\frac{f_n-f_{-n}}{2i}$, which may also be referred to as $f_n$ and $f_{-n}$ in our formalism.

\par All in all, $\mathbf C_{\mathcal N}$ is a vector space with $D_{\mathcal N}=(2\mathcal N+1)^D$ real dimensions $f_n$ each of which is labeled with an allowed Fourier mode $n$. We refer to elements of $\mathbf C_{\mathcal N}$ as \emph{$\mathcal N$-bounded functions}. Let us define the \emph{Lebesgue fractal measure} on $\mathbf C_{\mathcal N}$ by means of the fractal norm on each coordinate as:
\begin{equation}
d\ell_{(\kappa,\mu)}(f_n)=e^{\kappa (n^2+\mu^2)f_n^2/2}df_n.
\end{equation}

\noindent In principle, the \emph{Wiener fractal measure} comes out easily from this election for $\mathbf C_{\mathcal N}$ as
\begin{equation} \label {wiener3}
\begin{gathered}
dW_{\mathcal N}(t_N,\cdots,t_1)=\left(4\pi T \right)^{D_{\mathcal N}/2} ~\text{exp} \left( \frac{1}{4T} \sum_n \left[ \int_{f_n(-T)}^{f_n(T)} e^{\kappa (n^2+\mu^2)x^2/2} dx \right]^2  \right) \\
\times \left( \frac{1}{2\pi \tau_{N+1}} \right)^{D_{\mathcal N}/2} ~ \text{exp} \left(\sum_n \Bigg\{ -\frac{1}{2 \tau_{N+1}} \left[ \int_{f_n(t_N)}^{f_n(T)} e^{\kappa (n^2+\mu^2)x^2/2} dx \right]^2  \Bigg\} \right) \times \\
 \prod_{i=1}^N \Bigg\{  \left( \frac{1}{2\pi \tau_i } \right)^{D_{\mathcal N}/2} \text{exp} \left(\sum_n \Bigg\{ -\frac{1}{2 \tau_i} \left[ \int_{f_n(t_{i-1})}^{f_n(t_i)} e^{\kappa (n^2+\mu^2)x^2/2} dx \right]^2 + \frac{1}{2} \kappa (n^2+\mu^2)f_n^2(t_i) \Bigg\} \right) ~d^{D_{\mathcal N}} f_n(t_i) \Bigg\},
\end{gathered}
\end{equation} 

\noindent which essentially describes the Brownian motion of $\mathcal N$-bounded functions by referring to the fractal norm $\ell_{\kappa,\mu}$. Thus, the $\mathcal N$-bounded functions are subject to a Wiener process of fractality while they evolve stochastically in time. One should note that the restriction to $\mathbf C_{\mathcal N}$ is mandatory since integration is only well-defined for a finite-dimensional volume form. Hence, (\ref {wiener3}) is a well-defined probability measure as we promised before. However, since $\mathbf C = \bigcup_{\mathcal N} \mathbf C_{\mathcal N}$, then the path-integral will cover the whole of $\mathbf C$ as $\mathcal N \to \infty$.

\par Considering $\mathbf C$ as the Hilbert space of quantum states on $S$, which is commonly identified with $L^2(S)$, the Wiener fractal measure (\ref{wiener3}) is, in fact, the probability measure that describes the Brownian motion of the quantum states via an infinite-dimensional Wiener stochastic process. In the following, we will establish that this probability measure turns into the well-known Feynman's path-integral formulation of scalar quantum field theoreis. In principle, we will demonstrate that the Brownian motion of quantum states, in the sense of quantum mechanics, will produce the dynamics of quantum fields in the framework of quantum field theory.

\par Given a Wiener convergent measurable function $F(x_n(t_N),\cdots,x_n(t_1))$ the following integral exists and is well-defined for any $\mathcal N$;
\begin{equation} \label {wiener4}
I_{\mathcal N}:=\int_{U^{\mathcal N}_N}\cdots \int_{U^{\mathcal N}_1} F_{\mathcal N}(f_n(t_N),\cdots,f_n(t_1)) dW_{\mathcal N}(t_N,\cdots,t_1)
\end{equation}

\noindent wherein $F_{\mathcal N}(f_n(t_N),\cdots,f_n(t_1))$ is the restriction of $F(f_n(t_N),\cdots,f_n(t_1))$ to $\mathbf C_{\mathcal N}$, and $U^{\mathcal N}_i=U_i \bigcap \mathbf C_{\mathcal N}$. Actually, we are mostly interested in the behavior of the sequence $\{ I_{\mathcal N} \}_{\mathcal N=1}^{\infty}$ as $\mathcal N \to \infty$. It is worth to note that through this viewpoint the Wiener convergence condition must be replaced by $I_{\mathcal N}$ being a convergent sequence as $\mathcal N \to \infty$. A Wiener-integrable function $F(f_n(t_N),\cdots,f_n(t_1))$ which provides a convergent sequence $I_{\mathcal N}\to I$ is called a \emph{Wiener renormalizable} function. The well-defined value $I$ is also referred to as the \emph{path-integral of $F(f_n(t_N),\cdots,f_n(t_1))$}. 

\par It is easy to see that if $F(f_n(t_N),\cdots,f_n(t_1))=e^{V_N(f_n(t_N))} \cdots e^{V_1(f_n(t_1))}$ is bounded to one from above, which happens for negative potentials $V_i(f_n(t_i))$, then $\{I_{\mathcal N}\}$ would be a decreasing sequence, hence converges to some definite value $I\in \mathbb R$.\footnote{Obviously, if $V_i(f_n(t_i)) \leq 0$, then integration on more Fourier modes with the Wiener fractal measure will lead to a less amount, hence $I_{\mathcal N}$ is decreasing and convergent.} This condition is definitely equivalent to the set of energy eigenvalues being bounded from below, which is mandatory to gain a well-defined quantum field theory. However, as we will see in the following, any local potential $V(f_n)=V_i(f_n)$ expressed as an integral over the space manifold $S$ leads to an integrable function $F(f_n(t_N),\cdots,f_n(t_1))$ whenever the interaction term of $V(f_n)$ belongs to a renormalizable scalar quantum field theory. Therefore, all the local renormalizable interaction terms of scalar quantum field theories have intimate correlations to Wiener renormalizable functions.\footnote{As we will see in the following there is a slight difference between Wiener renormalizability and the renormalizability of a quantum field theory, while actually, the former class contains the latter one as a proper subclass.}

\par In the following, we will see that the Wiener process of fractality provides a mathematically well-defined formulation of Feynman's path-integral approach in quantum field theory. This is done by transferring to a complex measure from the Wiener fractal measure (\ref {wiener3}) via an approximation mechanism which is necessary to obtain a local formalism. Also, to control the possible divergencies which may emerge because of removing the non-local terms from the exponent of the Wiener fractal measure an infinitesimal heat kernel is imposed into the fractal mass which yields the causality of the theory. Actually, (\ref {wiener3}) and (\ref {wiener4}) provide the most consistent framework of the path-integral formulation of scalar quantum field theories. The approximated version of (\ref {wiener3}) will lead to two interesting achievements for relativistic quantum field theories which are the main subjects of the next two sections. 


\par
\section{Emergence of Lorentz Symmetry}
\setcounter{equation} {0}
\par 

\par According to the involved non-local terms calculating the Wiener path-integral (\ref {wiener4}) is a highly complicated process, unless we consider some appropriate approximations in the exponent of the Wiener fractal measure. However, we should emphasize that the only correct and exact measure that must be employed to extract the Green's functions of a quantum field theory is the original formula of (\ref {wiener3}), hence any approximation of the Wiener fractal measure will lead to some approximated formulas for expectation values of the quantum field theory. In principle, we are interested in a local approxmated version of $dW_{\mathcal N}(t_N,\cdots,t_1)$ which is actually derived for two simplifying assumptions as following:

\par \textbf{i)} Ignore the contributions at high amounts of $f_n$s, which upon \textbf{Theorem 1} and \textbf{Theorem 2} it means to exclude quantum fields with high amount of fractality. In principle, it is equivalent to suppose each $U^{\mathcal N}_i\subset \mathbb R^{D_{\mathcal N}}$ in (\ref {wiener4}) is a compact set. However, $U^{\mathcal N}_i$ will be replaced by $\mathbb R^{D_{\mathcal N}}$ at the end of the approximation process.

\par \textbf{ii)} Set $N \to \infty$, which leads to: $\tau_i=\theta \to 0$, $1\leq i \leq N$.\\

\par With these two assumptions we have;
\begin{equation} \label {cece}
\begin{gathered}
\Bigg\{ \int_{f_n(t_{i-1})}^{f_n(t_i)} e^{\kappa (n^2+\mu^2)x^2/2} dx \Bigg\}/\tau_i = \partial_tf_n(t_i) e^{\kappa (n^2+\mu^2)f_n^2(t_i)/2} \\ =\partial_tf_n(t_i)
+ \partial_tf_n(t_i) \Big\{ \kappa(n^2+ \mu^2) f_n^2(t_i)  \Big\} + \frac{1}{2} \partial_tf_n(t_i) \Big\{  \kappa(n^2+ \mu^2) f_n^2(t_i)  \Big\}^2 + \cdots
\approx \partial_tf_n(t_i),
\end{gathered}
\end{equation}

\noindent provided $\kappa f_n^2 \ll1$. Thus (\ref {wiener3}) turns into:
\begin{equation} \label {wiener5}
\begin{gathered}
dW_{\mathcal N}(t_N,\cdots,t_1)\\
\approx \text{exp} \left(\sum_n \Bigg\{ -\frac{1}{2} \left(\partial_t f_n(t_i)\right)^2 \tau_i + \frac{1}{2} \kappa (n^2+\mu^2)f_n^2(t_i) \Bigg\} \right) \times C_I^F \times \prod_{i=1}^N \left(  \left(  \frac{1}{\sqrt{2\pi \theta}} \right)^{D_{\mathcal N}} d^{D_{\mathcal N}} f_n(t_i) \right) \\
\approx  \text{exp} \left(\frac{c}{\frak h} \int_X \Bigg\{ -\frac{1}{2 c^2} \left(\partial_t f(t,x)\right)^2 + \frac{1}{2} \sum_{i=1}^D \left( \partial_i f(t,x) \right)^2 +\frac{1}{2} \frac{\frak m^2c^2}{{\frak h}^2} f^2(t,x) \Bigg\} d^Dx dt\right) ~\frak D f~~~~~~~~~~~~~\\
= \text{exp} \left(\frac{1}{\frak h} \int_X \Bigg\{- \frac{1}{2} \eta^{\mu \nu} \partial_\mu f(t,x) \partial_\nu f(t,x) + \frac{1}{2}  \frac{\frak m^2c^2}{{\frak h}^2}  f^2(t,x) \Bigg\} d^Dx dx^0 \right) \frak D f, ~~~~~~~~~~~~~~~~~~~~~~~~~~~
\end{gathered}
\end{equation} 

\noindent wherein we have set:
\begin{equation} \label {weiner6}
\frak m=\frac{\pi^3 \mu \theta  L^{D-3} }{\kappa},~~~~~~~~~~~~c=\frac{L\sqrt{\kappa}}{\pi \sqrt{\theta}  },~~~~~~~~~~~~\frak h=\frac{\pi  \sqrt{\theta}L^{D-1}}{\sqrt{\kappa}},~~~~~~~~~~~X=[-T,T]\times S,
\end{equation}

\noindent while the volume form $\frak D f$ is:
$$\frak D f=C_I^F \left( \frac{1}{2\pi \theta} \right)^{ND_{\mathcal N}/2}  \prod_{i=1}^N d^{D_{\mathcal N}} f_n(t_i),$$
\noindent for the overall (initial-to-final) normalization factor
$$C_I^F=\left( 4\pi T \right)^{D_\mathcal N/2} ~\text{exp} \left( \sum_n [ \int_{f_n(-T)}^{f_n(T)} e^{\kappa (n^2+\mu^2)x^2/2} dx ]^2  /4T \right).$$
\noindent Moreover, in the last line of (\ref {wiener5}) use has been made of standard notations $dx^0=c dt =c \theta$ and $\eta^{\mu \nu}=\text{diag}(1,-1,\cdots,-1)$, the Lorentz metric in $D+1$ dimensions. To transfer to physical scales the dimension of $f$, i.e. $T^{\frac{1}{2}} $, must be changed to $M^{\frac{1}{2}}T^{-\frac{1}{2}} L^{\frac{1}{2}(3-D)}$. Let $\gamma$ be such an scaling factor of dimension $M^{\frac{1}{2}}T^{-1} L^{\frac{1}{2}(3-D)}$ and define $\phi=\gamma f$. Thus, with this redefinition the Wiener fractal measure turns into:
\begin{equation} \label {wiener5'}
\begin{gathered}
dW_{\mathcal N}(t_N,\cdots,t_1)\\
= \text{exp} \left(\frac{1}{\hbar} \int_X \Bigg\{ -\frac{1}{2} \eta^{\mu \nu} \partial_\mu \phi(t,x) \partial_\nu \phi(t,x) + \frac{1}{2}  \frac{m^2c^2}{{\hbar}^2}  \phi^2(t,x) \Bigg\} d^{D+1}x \right) \frak D \phi,
\end{gathered}
\end{equation} 

\noindent for the physical mass $m=\frak m\gamma^2$ and the Dirac constant $\hbar =\frak h\gamma^2$, while we have simply considered
$$\frak D \phi =C_I^F \frac{1 }{ \left( 2\pi \theta \gamma^2  \right)^{ND_{\mathcal N}/2}   } \prod_{i=1}^N d^{D_{\mathcal N}} \phi_n(t_i) $$
\noindent as the \emph{Feynman measure}. In fact, if $D=3$ and $L , T \to \infty$, then the Wiener fractal measure becomes:
\begin{equation} \label {k-g}
dW_{\mathcal N}(t_N,\cdots,t_1)=  \text{exp} \left( - \frac{1}{\hbar} \int_{\mathbb R^4} \mathcal L_{K-G} ~d^4x \right) \frak D \phi
\end{equation}

\noindent for $\mathcal L_{K-G}$ the Klein-Gordon Lagrangian density.

\par Therefore, the Wiener fractal measure admits a natural Lorentz symmetry which stems from the substantial fractality of the scalar quantum fields. More precisely, the different signs of the time and the spatial dimensions within the Lorentz metric are the immediate implications of employing the fractality in the Wiener stochastic process. The sign of the time dimension comes from the Gaussian distribution in the Wiener's formula, but the signs of the spatial dimensions are the consequence of the fractal norm. Moreover, the partial derivations in (\ref {wiener5'}) are absolutely well-defined and justifiable, since as far as we concentrate on ${\mathcal N}$-bounded functions the included quantum fields in the Wiener fractal measure are smooth. Therefore, although the process of $\mathcal N \to \infty$ covers the whole space of fractal fields, according to the definition of the path-integral via (\ref {wiener4}), the smoothness of quantum fields is involved within the entire framework of the path-integration.

\par Obviously, any scalar quantum field theory could be similarly formulated by path-integration of some Wiener renormalizable function $F(x_n(t_N),\cdots,x_n(t_1))=e^{V(x_n(t_N))} \cdots e^{V(x_n(t_1))}$, by employing the approximated Wiener fractal measure (\ref {wiener5'}). For example, for $\phi^4$-theory in $\mathbb R^4$, we have; $V(x_n)\propto -\sum_{n,m} \left( x_m x_{n-m} \right)^2\leq 0$.\footnote{One should note that the leading term of the non-local part of the Wiener fractal measure is proportional to $f_n^4$ provided the time interval $\theta$ does not approach zero. This non-local term is easily verified as equivalent to $\left(\phi \star \phi \right)^2$ for $\star$ the convolution product. This is the most compatible non-local version for $\phi^4$-theory which automatically emerges in the Wiener fractal measure. We will reconsider this point in \cite{varshovi}.} Consequently, as we discussed above this interaction term leads to a Wiener renormalizable function with path-integral
\begin{equation} \label {phi-4}
\int \text{exp} \left(- \frac{1}{\hbar} \int_{\mathbb R^4} \mathcal  L_{\phi^4} ~d^4x \right) \frak D \phi := \int_{U_N^{\mathcal N}=\mathbb R^{D_{\mathcal N}}} \cdots \int_{U_1^{\mathcal N}=\mathbb R^{D_{\mathcal N}}} \text{exp} \left(- \frac{1}{\hbar} \int_{\mathbb R^4} \mathcal  L_{\phi^4} ~d^4x \right) \frak D \phi
\end{equation}

\noindent wherein $\mathcal L_{\phi^4}=\mathcal L_{K-G}-\frac{\lambda \hbar^{-1}}{4!}\phi^4$ is the Lagrangian density of $\phi^4$-theory for some positive constant $\lambda$.\footnote{As stated above $\lambda$ has to be a positive number since the Wiener path-integral (\ref {phi-4}) must lead to a well-defined amplitude as $\mathcal N \to \infty$. This is equivalent to the stability of the corresponding $\phi^4$-theory. See \cite{zee} for more discussions about unstable vacuum state of $\phi^4$-theory for negative $\lambda$.} Actually, despite assumption \textbf{(i)}, here we considered the intermediate integral domains $U^{\mathcal N}_i$s to be the whole space of $\mathcal N$-bounded functions $\mathbf C_{\mathcal N}$. Roughly speaking, we have considered that the contributions of highly fractal quantum fields could be ignored concretely.

\par Upon the above assumptions, the Wiener path-integral (\ref {phi-4}) is, in principle, a theoretical criterion for evaluating the probability of transferring the initial state $\phi_I$, given by Fourier coefficients $f_n(-T)$s, to the final state $\phi_F$, expanded by $f_n(T)$, in the target phase space of scalar functions or, in fact, the infinite-dimensional Hilbert space of quantum states. However, if the initial and the final states are set to be the vacuum field as $f_n(T)=f_n(-T)=0$, for all $n$, and the potential exponential function $F(x_n(t_N),\cdots,x_n(t_1))$ is multiplied by a given functional $\mathcal O(x_n(t_N)) \times \cdots \times \mathcal O(x_n(t_1))$, then the corresponding Wiener path-integral would be
\begin{equation} \label {exp-vev}
\int \text{exp} \left(- \frac{1}{\hbar} \int_{\mathbb R^4} \mathcal  L_{\phi^4} ~d^4x \right) \mathcal O(\phi) \frak D \phi=\langle \mathcal O(\phi) \rangle_{\text{VEV}}
\end{equation}

\noindent wherein $\langle \mathcal O(\phi) \rangle_{\text{VEV}}$ is the vacuum expectation value of the local operator $\mathcal O(\phi)$. On the other hand, for the Fourier modes $n_1,\cdots,n_l,m_1,\cdots,m_k \in \mathbb Z^D$, and with Fourier coefficients $\phi_n(t)=\gamma f_n(t)$, the LSZ reduction formula shows that
\begin{equation} \label {LSZ}
\begin{gathered}
\int \text{exp} \left(- \frac{1}{\hbar} \int_{\mathbb R^4} \mathcal  L_{\phi^4} ~d^4x \right) \Big\{ \phi_{n_1}(t_N) \cdots \phi_{n_k}(t_N) \phi_{m_1}(t_1) \cdots \phi_{m_l} (t_1)\Big\} ~ \frak D \phi
\end{gathered}
\end{equation}

\noindent effectively describes the scattering matrix element ${_\text{out}}\langle n_1,\cdots ,n_k | m_1, \cdots, m_l  \rangle_{\text{in}}$ modulo some overall factors on the Fourier modes. The exact formula and the unitarity of this matrix are better understood via the complexification of the Wiener fractal measure (\ref {wiener5}) in the next section. We emphasize that here we are actually working with quantum fields in Heisenberg's picture. According to the analytic structure of the symmetric form of the Wiener measure, the emergent quantum field theory is essentially time reversal, while upon the construction of the Wiener fractal measure over the space of $\mathcal N$-bounded functions we see that the theory admits the parity and the charge conjugate symmetries too.\footnote{However, in \cite{varshovi E-H} we will see that gravity automatically emerges in the Wiener fractal measure as an entropic-geometric force while it intrinsically respects the time's arrow of the second law of thermodynamics.}

\par Before closing this section we may wish to have a new look at the physical constants that emerged via approximating the Wiener fractal measure in (\ref {weiner6}). As we see all these constants depend on either the parameters of the massive fractal norm, i.e. $\kappa$ and $\mu$, or the structure of space-time, i.e. the spatial size $L$ and the time slicing width $\theta$, which the latter is substantially imposed by the potential functional $F$. Actually, $\kappa$ is of dimension $T^{-1}$, hence plays the role of frequency and includes some fundamental information about the particle. In other words, $\hbar \kappa$ would provide an important criterion for the step of the energy levels of the particle in compact space $S$. Moreover, since $\kappa$ could be absorbed in the time dimension within the speed of light $c$, and $n^2$ is the Fourier transform of the Laplacian operator,\footnote{As the kinetic part of the Hamiltonian operator} the appearance of the time partial derivative $\partial_0^2$ in the Gaussian distribution merges the space and time dimensions in a Lorentz symmetric way, generating the space-time as a unit continuum via the approximation (\ref{cece}). However, it is worth noting that based on distinct natures of spatial and time dimensions in the Wiener process, we deliberately use the terminology of \emph{space-time}, but not \emph{spacetime}, for addressing the total product manifold $X =[-T,T]\times S$ which appeared in (\ref {wiener5}) and (\ref {weiner6}).

\par However, because of its dimension $\kappa$ could be compared to $1/\theta$, which in its turn is included in the interaction term due. Thus, the only constant that explicitly inherits the fractal structure of quantum fields through the spatial dimensions is the fractal mass $\mu$. As we saw above $\mu$ depends partly on the spatial extension of the fractal field and partly on the type of the self-similarity due. For instance for a localized Weierstrass function of $(a,b)$-self-similarity we have:
\begin{equation}
\mu \propto \sqrt{\frac{a(b-a)}{s(1-a)(b-a)}}\sqrt{\frac{L-\frak l}{L}}=\text{Fractality} \times \text{Spacial Extension}.
\end{equation}

\noindent This formula confirms our physical intuition since it shows that the fractality of a wave function would be an equivalent parameter for the physical quantity of energy distributed through the space. Thus, the total energy of a particle, which is encoded as the total fractality via the exponent of the fractal norm, i.e. $\mu^2 + n^2$, has a background amount corresponding to the type of the asymptotic self-similarity $(a,b)$ of the included wave functions that is given by the fractal mass $\mu$, together with the Laplacian term $n^2$ which reveals the frequency of its oscillation. This resembles the energy formula of a $D$-dimensional quantum harmonic oscillator $E=\hbar \omega(\frac{D}{2} +\sum_{i=1}^D n_i)$, which has a ground state with $E=\frac{D}{2}\hbar \omega$ comparable to the fractal mass term $\kappa \mu$, whereas its oscillation energy contribution $E_{osc}=\hbar \omega( \sum_{i=1}^D n_i)$ reminds the oscillation role in the fractal norm, i.e. $\kappa (\sum_{i=1}^D n_i^2)$.


\par
\section{Wiener Complex Fractal Measure and Quantum Field Theory}
\setcounter{equation} {0}
\par 

\par Actually, the Wiener fractal measure loses its consistency and finiteness after implementing the approximation mentioned in the last section. In fact, the negative term in the exponent of the Gaussian distribution in the original Wiener fractal measure (\ref {wiener3}) always exceeds the positive parts appearing in the exponent of the fractal norm, hence the Wiener fractal measure is effectively given in terms of some heat kernels. Therefore, the formulation in (\ref {wiener3}) is well-defined and bounded for any $f_n$. However, when one transfers from (\ref {wiener3}) to (\ref {wiener5}) via approximation (\ref {cece}) this consistency is lost. In principle, the replacement of
\begin{equation} \label {c1}
~~~~~~~~~~\int_{f_n(t_{i-1})}^{f_n(t_i)} e^{\kappa(n^2+\mu^2)x^2/2}dx ~~~~~~~~~~\mapsto ~~~~~~~~~~ f_n(t_i)-f_n(t_{i-1})~~~~~~~~~~(1\leq i \leq N)
\end{equation}

\noindent in the exponent of the Gaussian function via the mentioned approximation approach causes the Wiener fractal measure to lose its powerful damping mechanism and includes indefinite values, hence being ill-defined or inconsistent. To overcome this problem, we have to compensate for the lost dominated non-local heat kernel which was removed and replaced by a linear term through the approximation process (\ref {c1}). In fact, incorporating any heat kernel beside the approximated formula would not guarantee the convergence of the approximated Wiener fractal measure, unless the exponent term of the measure is turned into a pure imaginary function since in this case, any heat kernel with a small quadratic exponent will restore the lost consistency.

\par In other words, although the integration of $e^{-ax^2}$ on $\mathbb R$ is not well-defined for negative $a$, its complexified version $e^{iax^2}$ would always have a finite integral but at the cost of adding some extra infinitesimal quadratic component into the exponent. Indeed, the main idea for treating the pathology of the approximated Wiener fractal measure stems from the analytic continuation of the measure and employing an infinitesimal heat kernel term to control the value of the integral. In fact, the following equation is used to overcome the inconsistency problem of the approximated Wiener fractal measure:
\begin{equation} \label {2103'}
\Big|\int_{\mathbb R} e^{(ia-\varepsilon)x^2}dx \Big|=\sqrt {\frac{\pi}{\sqrt{a^2+\varepsilon^2}}}~~~~~~~~~~  \underrightarrow{\varepsilon \to 0^+}~~~~~~~~~~ \sqrt{\frac{\pi}{|a|}}=\int_{\mathbb R} e^{-|a|x^2}dx,
\end{equation}

\noindent or more generally:
\begin{equation} \label {2103}
\lim_{\varepsilon \to 0^+}\Big| \int_{\mathbb R} e^{(ia-\varepsilon)x^2}f(x) dx \Big|=\Big| \int _{\mathbb R} e^{-|a|x^2} f(x) dx \Big|,
\end{equation}

\noindent for any analytic function $f(x)=\sum_{n=0}^\infty c_n x^n:\mathbb R \to \mathbb R$. Therefore, in order to have a well-defined approximated Wiener fractal measure we should employ the following replacement;
\begin{equation} \label {c3}
\begin{gathered}
\text{exp} \left(-\frac{1}{\hbar} \int_{\mathbb R^{D+1}} \Bigg\{\frac{1}{2} \eta^{\mu \nu} \partial_\mu \phi(t,x) \partial_\nu \phi (t,x) - \frac{1}{2}  \frac{m^2c^2}{{\hbar}^2}  \phi^2(t,x) \Bigg\} d^{D+1}x \right)\\
\mapsto\\
\text{exp} \left(\frac{i}{\hbar} \int_{\mathbb R^{D+1}} \Bigg\{\frac{1}{2} \eta^{\mu \nu} \partial_\mu \phi(t,x) \partial_\nu \phi (t,x) - \frac{1}{2}  \frac{(m^2-i \varepsilon) c^2}{{\hbar}^2}  \phi^2(t,x) \Bigg\} d^{D+1}x \right).
\end{gathered}
\end{equation}

\noindent The produced Wiener fractal measure due to the complexification (\ref {c3}) is referred to as the \emph{Wiener complex fractal measure}.  

\par As we saw above $\varepsilon$ has a critical role in producing a well-defined measure. However, since the causality of quantum field theories is an immediate consequence of the $i\varepsilon$-term \cite{peskin}, we see that the replacement (\ref {c3}) successfully revives the natural causality of the Wiener path-integral. An augmented beneficial $i\varepsilon$-term also appears in demonstrating the Feynman propagators or expectation values of some local operators via the limit of $T \to (1-i\varepsilon)\infty$ or equivalently for setting $p_0 \mapsto p_0+i\varepsilon$ with $p_0$ the energy of virtual particles \cite{peskin}. This mechanism is actually an alternative procedure for the replacement of $m \mapsto m-i\varepsilon$ within the Lagrangian density.

\par More precisely, the $i\varepsilon$-term is an appropriate alternative for the non-local terms of the fractal norm in the exponent of the Wiener fractal measure. Generally, upon the above argument, it can be inferred that in all path-integral formulations of quantum physics the $i\varepsilon$-term plays the role of the removed non-local terms to control the finiteness of the approximated Wiener fractal measure.\footnote{See \cite{witten} for the role of $i\varepsilon$-term in string theory.} However, it should be emphasized that the $i\varepsilon$-tem will not obviously reproduce the whole non-local effects in the Brownian motion of quantum states. Therefore, the $i\varepsilon$-term is only an approximate remedy for the inconsistency problem mentioned above and we lose all non-local information of the theory via the approximation (\ref{cece}). On the other hand, according to the significance of the $i\varepsilon$-term in figuring out the structure of the vacuum state of a quantum field theory \cite{weinberg}, it seems that the vacuum state has an intimate correlation to the non-local contributions and the fractality of quantum fields.

\par The above machinery for working out the genuine form of Feynman's path-integral formulation of scalar field theory could be simply generalized to a quantum field theory of $\frak n$ independent quantum fields $\phi_{(i)}$, $1\leq i \leq \frak n$. In this case the well-defined complexified approximated Wiener fractal measure would be;
\begin{equation} \label {c4}
\text{exp} \left(\frac{i}{\hbar} \int_{\mathbb R^{D+1}} \sum_{j=1}^{\frak n} \Bigg\{\frac{1}{2} \eta^{\mu \nu} \partial_\mu \phi_{(j)}(t,x) \partial_\nu \phi_{(j)}(t,x) - \frac{1}{2}  \frac{(m^2-i \varepsilon) c^2}{{\hbar}^2}  \phi_{(j)}^2(t,x) \Bigg\} d^{D+1}x \right) \frak D \phi.
\end{equation}

\noindent Especially, this well-defined measure naturally leads to the Maxwell theory for $\frak n=3$ and $m=0$ in the Coulomb gauge; $\partial_i A_i=A_0=0$. Therefore, it seems that the Wiener fractal process could be similarly considered for gauge field theories via some appropriate modifications within the fractal norm and consequently the Lebesgue fractal measure.\footnote{This is the main subject of \cite{varshovi}.} Moreover, Feynman's path-integral formulation of quantum mechanics can be worked out via the machinery we used above to extract the path-integral formulation of scalar quantum field theories. This is the subject of \textbf{Appendix A}.

\par In particular, the main conclusion of the last two sections is that the Lorentz invariance of relativistic quantum field theories is an approximated symmetry that holds at the local approximation of the theories and stems from a natural feature of the fractality of wave functions, which emerges via the fractal norm. As we saw above, the fundamental assumption for extracting the symmetry is the classical separation of space and time as two distinct entities. This idea has two major similarities to the Horava-Lifschitz theory of quantum gravity \cite{horava}:\\

\par \textbf{i)} \emph{Spacial and time dimensions are non-equivalent (anisotropic) at high energy levels.}

\par \textbf{ii)} \emph{Space is actually defined via the time foliations of the space-time 4-manifold.}\\

\par Although Horava's theory is a candidate formulation for the quantum theory of gravity it has fundamental similarities to our derived formulas for scalar quantum field theory. Actually, both Horava's assumptions have intimate correlations to the original idea of the Wiener stochastic process. In addition, the approximation process (\ref {cece}) of the Wiener fractal measure ceases for high amounts of $n^2$ and must contain next to the leading terms in the action. Hence, at least one has to consider $n^2 f_n^2(t) (\partial_t f_n(t))^2$ within the Fourier transform of the Lagrangian density, which is in fact a non-local term. However, Horava's prescription is to employ local terms with higher spatial derivatives at the Lifshitz point $z=3$ and to impose the symmetry of foliation preserving diffeomorphism invariance on the theory.\footnote{In \cite{varshovi E-H}, where we work out the Einstein-Hilbert theory from the Wiener fractal measure, more similarities have been discussed and established between the Horava-Lifshits gravity and the Wiener Brownian process of fractals.}

\par Before closing this section we prefer to have a brief look at the Wick rotation process. Wick rotation as a reliable method for obtaining solutions in Minkowski space from the solutions in Euclidean space via replacing some imaginary variables with real ones is indeed an immediate reformulation of the complexification method (\ref{2103}) we implimented by employing an $i\varepsilon$-term in the heat kernel and the analytic continuation of the approximated Wiener fractal measure (\ref {wiener5}). This process could be regarded as an (almost) equivalence relation between Lorentzian and Euclidean signatures of the space-time metric.\footnote{Nevertheless, some physical phenomena reject this equivalence relation within some relativistic effects of quantum field theory. See \cite{sorkin} for a more detailed discussion of the problem.}

\par Actually, the original formulation of the Wiener fractal measure (\ref {wiener3}) may explain  why nature does not embrace the Euclidean signature intrinsically. In particular, based on the above argumentation there is no correct signature of the space-time continuum and all we obtain within local Lagrangian formulations of nature are approximated structures at low energy regimes of the Wiener Brownian process. However, after accomplishing the approximation procedure (\ref {cece}) and obtaining (\ref {wiener5}) it would be natural to employ the regularization method of (\ref {2103}), which would lead us to thermal quantum field theories and the functional methods in condensed matter \cite{peskin}.\footnote{See \cite{landsman} for an alternative approach to real-time formalisms in thermal field theory via Bogoliubov transformations, the so-called \emph{thermo field dynamics}.}


\par
\section{Wiener Fractal Measure and Renormalization}
\setcounter{equation} {0}
\par 

\par Now we are ready to have a new look at the renormalizability problem of quantum field theories. Actually, the Wiener fractal measure provides a profound understanding of renormalizability. To see this in a better way let us turn back to the original (real valued) Wiener fractal measure and Eq. (\ref {wiener4}). One should note that each individual integral of (\ref {wiener4}) for a local potential $V_i(f_n)$ is actually of the following form
\begin{equation} \label {con1}
\mathcal J_{\mathcal N}=\int_{\mathbb R^{D_{\mathcal N}} } \text{exp}\left( \Bigg\{ -\frac{1}{2\tau} \sum_n \Big|\int^{f_n}_{y_n} e^{\kappa (n^2+\mu^2) x^2/2} dx \Big|^2  \Bigg\}  + \int_S \Big\{\frac{\kappa}{2} \left( |\nabla f|^2+ \mu^2 f^2\right)+ V_i(f) \Big\}~d^Dx \right) d^{D_{\mathcal N}}f_n,
\end{equation}

\noindent for some $y_n \in \mathbb R$ and with $\nabla f$ the gradient of $f$. Therefore, $I_{\mathcal N}$ is roughly written as; $I_{\mathcal N} \propto \mathcal J_{\mathcal N}^N$. In fact, in physics we readily assume that the number of the time slices $N$ could be enlarged arbitrarily, hence we may simply replace the role of $N$ with $g( \mathcal N)$, for $g(\mathcal N):\mathbb N \to \mathbb N$ a monotone increasing function, and fix $t_{g(\mathcal N)}=T$ and $t_{-g(\mathcal N)}=-T$ for all $\mathcal N$. This assumption may cause a slight redefinition of the Wiener renormalizable functions, which is now referred to as the \emph{Wiener renormalizable sequence}. Actually, a Wiener renormalizable sequence is a set of functions $\{F_{\mathcal N}(f_n(t_{g(\mathcal N)}),\cdots,f_n(t_{-g(\mathcal N)})\}$, which has two properties;\\

\par \textbf{i)} Each $F_{\mathcal N}(f_n(t_{g(\mathcal N)}),\cdots,f_n(t_{-g(\mathcal N)}))$ is a Wiener fractal convergent function in $\mathbf C_{\mathcal N}$.

\par \textbf{ii)} The sequence of the Wiener path-integrals
\begin{equation} \label {badan 1}
I_{\mathcal N}=\int F_{\mathcal N}(f_n(t_{g(\mathcal N)}),\cdots,f_n(t_{-g(\mathcal N)})) ~dW_{\mathcal N}(t_{g(\mathcal N)}, \cdots, t_{-g(\mathcal N)})
\end{equation}

\noindent converges to some definite value $I \in \mathbb R$ as $\mathcal N \to \infty$, the so-called \emph{Wiener path-integral of $F_{\mathcal N}$s}.\\

\par Indeed, a renormalizable quantum field theory is a Wiener renormalizable sequence that fulfills three more conditions;\footnote{However, we should emphasize that these three properties are in fact fractal-geometric/stochastic interpretation of RG flow. For example, the property \textbf{(iv)} effectively provides an equivalent definition for effective field theories according to the RG flow.}\\

\par \textbf{iii)} The potential exponential functions $F_{\mathcal N}(f_n(t_{g(\mathcal N)}),\cdots,f_n(t_{-g(\mathcal N)}))$ could be written by means of a local potential $V_{\mathcal N}(f_n)$ within the time slicing $t_{-g(\mathcal N)}< \cdots < t_{g(\mathcal N)}$ as;
\begin{equation} \label {a}
F_{\mathcal N}(f_n(t_{g(\mathcal N)}),\cdots,f_n(t_{-g(t_{\mathcal N})})= \exp\left( {\int_S V_{\mathcal N}(f_n(t_{g(\mathcal N)}) ~d^Dx } \right) \cdots \exp \left({\int_SV_{\mathcal N}(f_n(t_{-g(\mathcal N)}) ~d^Dx }\right).
\end{equation}

\par \textbf{iv)} For each $\mathcal N$ the potential $V_{\mathcal N}(f_n)$ is a polynomial of $f_n$s of order $\geq 2$ which contains a fixed number of constants. The coefficient of $f_n^2$ in $V_{\mathcal N}(f_n)$ is effectively the \emph{mass renormalization} and the coefficients of the higher order terms are the \emph{running coupling constants.} The emergent coefficient $Z$ before the Gaussian term of the Wiener fractal measure is indeed the \emph{field strength renormalization.}

\par \textbf{v)} If $g(\mathcal N)-g(\mathcal N-1) \propto \frac{cT}{L}$, as $\mathcal N \to \infty$, then integrating out the highest frequencies and the intermediate time intervals causes the following \emph{renormalization group} process;\footnote{Actually, if $\frak L=\{- g(\mathcal N), \cdots, g(\mathcal N) \}$ is a lattice, then any real function $\frak f:L\to \mathbb R$ could be simply converted to a real function as $\tilde {\frak f}:[-T,T] \to \mathbb R$ via the inverse Fourier transformation as: $$\tilde {\frak f}(t)=\sum_{n=1}^{g(\mathcal N)} \frak f(n) \cos\left(\frac{n\pi t}{T}\right) + \sum_{n=-1}^{-g(\mathcal N)} \frak f(n) \sin\left(\frac{|n|\pi t}{T}\right) +\frac{1}{2} \frak f(0).$$ Hence, transferring from $\mathcal N$ to $\mathcal N-1$ will cause $\tilde{\frak f}(t)$ to lose $\Delta E=\left( g(\mathcal N)-g(\mathcal N-1)\right) \frac{\hbar \pi}{T}$ amount of energy. We use this idea for potential functions $F_{\mathcal N}\left(f_n(t_{g(\mathcal N)}),\cdots,f_n(t_{-g(t_{\mathcal N})}) \right)$ with $\Delta E \propto \frac{\hbar \pi}{L}$. This revives the Wilsonian process of integrating out the highest amount of energy included in a cutoff shell with fixed radius and thickness $\Delta E$ in the Euclidean momentum space defined by the Wick rotation.}
\begin{equation} \label {c}
~~~~~~~~~~F_{\mathcal N}(f_n(t_{g(\mathcal N)}),\cdots,f_n(t_{-g(\mathcal N)})) \stackrel{}{\longrightarrow} F_{\mathcal N-k}(f_n(t_{g(\mathcal N-k)}),\cdots,f_n(t_{-g(\mathcal N-k)})),~~~~~(k \in \mathbb N)
\end{equation}

\noindent as $\mathcal N\to \infty$, $T \to \infty$, and $ L \to \infty$. Through this process $b=\lim_{\mathcal N \to \infty} \frac{g(\mathcal N-1)}{g(\mathcal N)}$ effectively describes the RG flow \emph{scaling parameter}, and the survived polynomials in the limit of $b \to 0$ are the \emph{relevant operators}.\\

\par Actually, the above process in performing the Wiener path-integrals can be similarly converted to the Wiener complex fractal measure. Hence, the process of integrating out the highest frequencies and the intermediate time sections in \textbf{(v)} will naturally lead to the renormaliation group flow method for relativistic quantum field theories. It is worth to emphasize that both the Bogoliubov's \cite{bogo} and the Wilson's \cite{wilson1, wilson2} viewpoints to the RG flow machinery give rise to profound understandings of and guidelines towards the fractality of quantum fields.\footnote{See also \cite{zimm, shirkov1, shirkov2} for more discussion about Bogoliubov's special insights toward the RG flow for quantum fields.}

\par The original Wiener fractal measure (\ref {wiener3}) is a pure probability measure, hence its integrals would end up with finite values for appropriate renormalizable potentials terms. Therefore, the infinities appearing in loop calculations in interacting quantum field theories stem from removing the non-local terms of the exponent of the Gaussian distribution in the Wiener fractal measure (\ref {wiener3}) which we accomplished above to extract an approximated but local Lagrangian formulation for the theory of scalar quantum fields. In other words, the regularization and renormalization methods are indeed theoretical techniques to compensate for the fatal pathology of approximation (\ref {wiener5}). Therefore, handling the appeared non-local terms in the Wiener fractal measure, instead of utilizing various regularization methods and renormalization procedures would be an innovative theoretical trick to work out a mathematically well-defined path-integral formulation of quantum gravity. Indeed, this paper is the first step in practice for extracting such a promised quantum theory of gravity.

\par On the other hand, as we explained above the renormalization group flow of the Wiener path-integral (\ref {badan 1}) will cause the higher non-local terms to disappear (approximately) and then the Wiener path-integral would lead to the Feynman's path-integral formulation of an effective local field theory at low energy levels which contains only stable renormalizable local interactions \cite{peskin}. Thus, one may be optimistic that the Wiener path-integral of Brownian motion in infinite-dimensional Hilbert spaces equipped with the fractal norm could help us with better understanding the original theory for scalar quantum fields at high energy levels. 


\par
\section{Summery and Conclusions}
\setcounter{equation} {0}
\par 

\par This paper is, in fact, the first step of practice to figure out and explain the fundamental features of the path-integral formulation of quantum gravity within a consistent framework of Wiener's stochastic process which stands on a footing that consists of the fractality of RG flow renormalization and the Brownian motion in the infinite-dimensional Hilbert space of quantum states. In this present perusal, we are essentially concerned about the genuine formula of the Wiener measure for infinite-dimensional Brownian motion of scalar functions as quantum states on a flat space-time manifold. The theoretical approach is based on two principal aspects of quantum physics:\\

\par \textbf{1) Heisenberg's Uncertainty Principle and the Stochasticity of the Dynamics:} The stochastic propagation of quantum fields due to Heisenberg's uncertainty principle is the most essential feature of quantum physics that has been encoded within Wiener's path-integral measure.\\

\par \textbf{2) Wilsonian Renormalization of High Energy Contributions and Fractality:} Employing the background fractal geometry of the Wilsonian renormalization group flow in studying the infinite-dimensional Brownian motion and extracting the genuine formula of Feynman's path-integral measure.\\

\par We combined these two basic foundations of quantum physics by allowing the quantum states to evolve stochastically via an infinite-dimensional Wiener process with a norm of fractality. In the first step, we considered Hardy's nowhere differentiability condition for Weierstrass-like fractal functions and obtained a non-linear differential equation for the asymptotic behavior of Fourier-Laplace coefficients due. Finding the solution of the differential equation, we obtained an exponentially increasing norm on the phase space, the so-called \emph{fractal norm}, which transfers the fractal functions to the extremely far regions of the Hilbert space. Then, we employed the Wiener measure of the fractal norm, known as the \emph{Wiener fractal measure}, to study the Brownian motion of the quantum states on a flat space-time by considering the involved fractality. We established that this fractal geometry-based infinite-dimensional Brownian motion automatically leads to a number of interestin achievements:\\

\par \textbf{a)} \emph{The Wiener measure shows up an intrinsic Lorentz symmetry at its first local approximation}.\\

\par Hence, it is established that Lorentz invariance could be regarded as an approximate symmetry of local interactions of nature, as has already been shown in some aspects of string theory \cite{horava}. This conclusion is obtained without any augmentation of mathematical tools that have been imposed on the Wiener measure by hand, such as the analytic continuation of the measure, incorporating an imaginary mass,\footnote{These two methods cease the Wiener measure to keep its consistency as a well-defined probability measure.} regularizing the measure, and geometric approximations, which all have already been done before.\footnote{See for example \cite{bodmann, dau, dau2, dau3, and, nelson, cameron1, cameron2, chaichian}. For another viewpoint on Feynman's path-integral formulation see \cite{ito1, ito2, gelfand, kac}. The readers are also referred to \cite{klauder} for a historical review on path-integral and its various stochastic formulation via the Wiener measure.} The next astonishing outcome of the above arguments is:\\

\par \textbf{b)} \emph{The first local approximation of the Wiener fractal measure gives rise to the scalar field theory as the admitted theory of Higgs boson in the Standard Model}.\\

\noindent This leads to a probabilistic formalism of scalar quantum field theories within two formalisms in real analysis (including non-local terms that retain and guarantee the finiteness of the theory) and complex analysis (that consists of only local terms that are accompanied by an $i\varepsilon$-mass term for consistency). Although, this paper aims to produce a reformulation of Feynma's path-integral measure, one may conclude that all path-integral formulations of quantum physics could be figured out and reformulated via the Wiener stochastic process of the quantum states within the underlying fractal geometry. Therefore, it seems that the original form of the Wiener fractal measure can be regarded as a reliable framework to formulate the fundamental interactions of nature including gravity (as we have already established in \cite{varshovi E-H}) and Maxwell's and Yang-Mills theories (which is accomplished in \cite{varshovi}).

\par The distinguishing feature of this research concerning previous similar works that aimed to find a correlation between Feynman's path-integral measure and the Wiener measure is that here we do not aim to reproduce Feynman's path-integral formulation by improving the Wiener's standard formula but to find the correct Feynman's path-integral measure by renormalizing the Wiener's measure in infinite-dimensional Hilbert spaces. The next difference is that our method has been entirely independent of any doubtful assumptions related to the mathematics of the Wiener's formula, and the whole results have been obtained from a pure fractal geometric basis and the stochastic process. Hence, one may be optimistic that the achieved results of this perusal may help us to perceive and obtain more profound understandings and intuitions about the foundations of quantum physics and fundamental interactions of nature.


\section{Acknowledgments}

\par It should be noted that this research was in part supported by a grant from IPM (No. 1402830035). 


\appendix

\section{\\Quantum Mechanics Revisited via the Wiener Fractal Measure}

\par In this appendix, we will prove that the Wiener fractal measure can also produce the Feynman's path-integral formulation of quantum mechanics within a concrete procedure. Indeed, the Feynman's path-integral formulation of quantum mechanics must be understood via the complexification (\ref {2103}), whereas we do not need the symmetry (\ref{wiener1}). To work out the Feynman's path-integral formulation of quantum mechanics the Wiener complex fractal measure is restricted to the space of $\mathbf C_{\mathcal N_0}$, for some fixed $\mathcal N_0$, so that the whole reserved energy in $\mathbf C_{\mathcal N_0}$ is of order $mc^2$, for $m$ the rest mass of the quantum particle. Hence:
\begin{equation} \label {mass-C}
\sum_n \frac{\pi^2\hbar^2 n^2}{L^2}=\left( \frac{2D \pi^2 \hbar^2}{L^2}\right) \frac{\mathcal N_0(\mathcal N_0+1)(2\mathcal N_0+1)}{6} \propto mc^2,
\end{equation}

\noindent wherein the summation is over all Fourier modes $n$ which belong to $\mathbf C_{\mathcal N_0}$. Actually, one may neglect the high energy contributions. In principle, in quantum field theory, we are mostly concerned about the asymptotic behavior of the Wiener path-integrals $I_{\mathcal N}$ introduced in (\ref {wiener4}) as $\mathcal N \to \infty$, hence employing a renormalization approach is mandatory. But, here, there are no such asymptotic behaviors, and therefore, no renormalization is needed for any potential function.

\par To work out the Feynman's quantum mechanical path-integral we have to consider a semi-classical decomposition. To see this we need some simple definitions. Let $D^{x_0}(x)$, for $x_0 \in S$, be the restriction of the Dirac delta function $\delta\left((x-x_0)/L\right)$ to $\mathbf C_{\mathcal N_0}$ multiplied by $\gamma \sqrt{\theta}$, which is inserted for adjusting the physical dimensions. Set $D_{\mathcal N_0}:=\{D^{x_0}(x)|x_0 \in S\}$, the so-called \emph{$\mathcal N_0$-bounded Dirac space}. Actually, $D_{\mathcal N_0}$ is a $D$-dimensional subspace of $\mathbf C_{\mathcal N_0}$ with Fourier coefficient;\footnote{Note that here we have no summation rule on similar indices inside the arguments of trigonometric functions.}.
\begin{equation} \label {D_n}
{D^{x_0}}_n=\gamma \sqrt{\theta} \left( \prod_{n_i=0} \frac{1}{2} \times \prod_{n_i > 0} \cos\left(n_i\pi x^i_0/L\right) \times \prod_{n_i<0} \sin\left(|n_i|\pi x^i_0/L\right) \right).
\end{equation}

\noindent for the Fourier mode $n=(n_1,\cdots,n_D)$. Let us get more information about the $\mathcal N_0$-bounded Dirac space. Set $A_n:=\{n'=(n'_1,\cdots,n'_D)| n'_i=\pm n_i\}$ with $n=(n_1,\cdots,n_D)$ which fulfills $0\leq n_i \leq \mathcal N_0$ for all $1\leq i \leq D$. Then, we see that:
\begin{equation} \label {fariborz}
\sum_{n' \in A_n} { D^{x_0}_{n'} }^2=\frac{\gamma^2 \theta}{4^{z_n}}, 
\end{equation}

\noindent where $z_n$ is the number of zeros in $n$. Therefore, we obtain:
\begin{equation} \label {mushkhushu}
\sum_n { D^{x_0}_n }^2=\gamma^2 \theta \left(\frac{1}{4}+\mathcal N_0  \right)^D.
\end{equation}

\noindent Thus, the $\mathcal N_0$-bounded Dirac space is a $D$-dimensional subspace of $\mathbf C_{\mathcal N_0}$, included in the sphere of radius
\begin{equation} \label {sargon}
R_{\mathcal N_0}=\gamma \sqrt {\theta \left(\frac{1}{4}+\mathcal N_0  \right)^D }.
\end{equation}

\noindent Hence, the $\mathcal N_0$-bounded Dirac space is a compact subset of $\mathbf C_{\mathcal N_0}$.

\par Now, let $D_{\mathcal N_0}^{\perp}$ be the supplementary or perpendicular dimensions to $D_{\mathcal N_0}$ in $\mathbf C_{\mathcal N_0}$. The Wiener fractal measure is decomposed as:
\begin{equation} \label {dec-wien}
dW_{\mathcal N_0}=dW_{D_{\mathcal N_0}} \times dW^{\perp}.
\end{equation}

\noindent Moreover, according to (\ref{fariborz}) for each $D^{x_0}(x)$ and any Fourier mode $n=(n_1,\cdots,n_D)$ we have :
\begin{equation} \label {lond}
\sum_{n' \in A_n} ( \mu^2 + {n'}^2 ) \left( D^{x_0}_{n'} \right)^2 =( \mu^2 + n^2 ) \frac{\gamma^2 \theta}{4^{z_n}}.
\end{equation}

\noindent Thus, the exponent of the fractal norm is a constant number for all element of $D_{\mathcal N_0}$ and does not contribute to integrating with the relevant Wiener fractal measure $dW_{D_{\mathcal N_0}}$. On the other hand, since $D_{\mathcal N_0}$ is a $D$-dimensional space we simply read; $dW_{D_{\mathcal N_0}}  \propto \frak Dx$, for $\frak Dx :=\prod \frac{1}{\left( 2\pi \theta \right)^{D/2}} d^Dx$, the standard Feynman's path-integral measure for quantum mechanics. Therefore, the complexified approximated form of the Wiener fractal measure via the decomposition (\ref {dec-wien}) is;
\begin{equation} \label {w-dec}
\begin{gathered}
dW_{D_{\mathcal N_0}} \times dW^{\perp}=\exp \left( \frac{iL^D}{\hbar} \int_0^T  \sum_n \Bigg\{ \frac{1}{2c}  \dot{D^{x}_n}^2 +\dot{ D^{x}_n}\partial_0 \phi_n(t) + i \varepsilon \frac{ c^2}{\hbar^2} {D_n^x}^2    \Bigg\} dt \right) \frak D x\\
\times~ \text{exp} \left(\frac{i}{\hbar} \int_{X} \Bigg\{\frac{1}{2} \eta^{\mu \nu} \partial_\mu \phi(t,x) \partial_\nu \phi (t,x) - \frac{1}{2}  \frac{(m^2-i \varepsilon) c^2}{{\hbar}^2}  \phi^2(t,x)  \Bigg\} d^{D+1}x \right) \frak D \phi^{\perp},
\end{gathered}
\end{equation}

\noindent for $X=[0,T]\times S$. However, the mixed term $\dot{D^{x}_n}\partial_0 \phi_n(t)$ is odd for variables $\phi_n$s, hence cancels out from the measure due to orthogonality of $D_{\mathcal N_0}$ and $D_{\mathcal N_0}^\perp$. Moreover, via a similar reasoning we explained above one can see that:
\begin{equation} \label {w-dec2}
L^D \sum_n \dot{D^x_n}^2=P_{\mathcal N_0} \sum_{i=1}^D  \dot{x^i}^2 + \text{mixed term}.
\end{equation}

\noindent with;
\begin{equation} \label {mass-P}
P_{\mathcal N_0}=\frac{\hbar \kappa}{c} \left( \frac{\mathcal N_0(\mathcal N_0+1)(2 \mathcal N_0+1)}{6} \right) \left(\frac{5}{4} \right)^{D-1}.
\end{equation}

\noindent The mixed term consists of $\dot{x^i}\dot{x^j}$, $1\leq i\ne j\leq D$, which is odd for each variable $x^i$, hence cancels out at the first approximation via the path-integration. Therefore, based on (\ref{mushkhushu}) and (\ref{sargon}) we obtain;
\begin{equation}  \label {w-dec1}
dW_{D_{\mathcal N_0}} \approx \exp \left( \frac{i}{\hbar} \int_0^T  \Bigg\{ \frac{P_{\mathcal N_0} }{2c} \sum_{i=1}^D \dot{x^i}^2+i \varepsilon \frac{R_{\mathcal N_0}^2 L^D c^3}{\hbar^2} \Bigg\} dt \right) \frak D x.
\end{equation}

\noindent In fact, $P_{\mathcal N_0}/c$ is of the mass dimension, comparable to the particle's mass $m$ computed in (\ref {mass-C}). By adjusting $\kappa$ we can simply write: $P_{\mathcal N_0}/c=m$. Indeed, based on (\ref {mass-C}) and (\ref{mass-P}) if we adjust $\kappa$ once then for all quantum particles with different masses we obtain $P_{\mathcal N_0}/c=m$. In addition, the $\varepsilon$-term in $dW_{D_{\mathcal N_0}}$ is also a constant and does not participate in the process of path-integration. Thus, for any quantum particle with mass $m$ and at low energy levels we obtain;
\begin{equation} \label {c5}
dW_{D_{\mathcal N_0}}  \approx \text{exp} \left(  \frac{i}{\hbar} \int_0^T \Bigg\{ \frac{ m}{2}  \left(\frac{d x}{dt}\right)^2 \Bigg\} dt \right) \frak D x.
\end{equation}

\noindent If the quantum particle is subject to some external force given by the local potential $V(x)+V(\phi^\perp)$ the Wiener path-integral would be;
\begin{equation}
I \approx \int \text{exp} \left(  \frac{i}{\hbar} \int_0^T \Bigg\{ \frac{ m}{2}  \left(\frac{d x}{dt}\right)^2 -V(x) \Bigg\} dt \right) \frak D x \times  \int \exp \left(- \frac{i}{\hbar} \int_{X} V(\phi^\perp) d^4x \right)dW^\perp,
\end{equation}

\noindent which leads to the celebrated Feynman's path-integral formula for quantum mechanics \cite{feynman, feynman2};
\begin{equation}
I \propto \int \text{exp} \left(  \frac{i}{\hbar} \int_0^T L ~ dt \right) \frak D x,
\end{equation}

\noindent wherein $L$ is the classical Lagrangian and the integration of the measure $\frak D x$ is taken over all paths from the initial point to the final destination.


\end{document}